\documentclass[aip,jcp,showpacs,amsmath,amssymb,amsfonts,superscriptaddress,reprint]{revtex4-1}

\usepackage{graphicx}
\usepackage{dcolumn}
\usepackage{bm}
\usepackage{epsf}
\usepackage{color}
\usepackage[colorlinks=true,citecolor=blue,linkcolor=blue,urlcolor=blue]{hyperref}

\newcommand{\la}{\left\langle}
\newcommand{\ra}{\right\rangle}
\newcommand{\pd}{\partial}
\newcommand{\de}[1]{\delta\left(#1\right)}
\newcommand{\td}{\mathrm{d}}

\newcommand{\e}[1]{\exp{\left(#1\right)}}
\newcommand{\lo}[1]{\ln{\left(#1\right)}}

\newcommand{\bla}{bla\\bla\\bla\\bla\\bla}

\newcommand{\PRE}{Phys. Rev. E}
\newcommand{\PRL}{Phys. Rev. Lett.}
\newcommand{\PRX}{Phys. Rev. X}
\newcommand{\EPL}{EPL (Europhys. Lett.)}
\newcommand{\RMP}{Rev. Mod. Phys.}

\newcommand{\mb}[1]{\mbox{\boldmath$#1$}}
\newcommand{\mc}[1]{\mathcal{#1}}

\newcommand{\mrm}[1]{\mathrm{#1}}

\begin{document}

\title{Optimal driving of isothermal processes close to equilibrium}

\author{Marcus V. S. Bonan\c{c}a}
\email[]{mbonanca@ifi.unicamp.br}
\affiliation{Instituto de F\'isica 'Gleb Wataghin', Universidade Estadual de Campinas, 13083-859, Campinas, S\~{a}o Paulo, Brazil}
\altaffiliation{on leave of absence from Universidade Estadual de Campinas}
\affiliation{Department of Chemistry and Biochemistry and Institute of Physical Sciences and Technology, University of Maryland, College Park, Maryland 20742, USA}
\author{Sebastian Deffner}
\email[]{sebastian.deffner@gmail.com}
\affiliation{Department of Chemistry and Biochemistry and Institute of Physical Sciences and Technology, University of Maryland, College Park, Maryland 20742, USA}

\date{\today}

\begin{abstract}
We investigate how to minimize the work dissipated during nonequilibrium processes. To this end, we employ methods from linear response theory to describe slowly varying processes, i.e., processes operating within the linear regime around quasistatic driving. As a main result we find that the irreversible work can be written as a functional that depends only on the correlation time and the fluctuations of the generalized force conjugated to the driving parameter. To deepen the physical insight of our approach we discuss various self-consistent expressions for the response function, and derive the correlation time in closed form. Finally, our findings are illustrated with several analytically solvable examples. 
\end{abstract}


\maketitle

\section{Introduction \label{sec:intro}}

All physical devices operate in finite time, and, hence, inevitably dissipate energy. This observation is what lies beneath the various formulations of the second law of thermodynamics. A particular elucidating statement of this \textit{law} is the maximum work theorem, that predicts that the maximally extractable work during isothermal processes is given by the free energy difference $\Delta F$ \cite{callen_85}. Thus, the amount of energy that is lost during any real, physical process, i.e., the work dissipated into the environment is given by $W_\mrm{irr}\equiv W-\Delta F$, where $W=\int\td\mc{W}\, \mc{P}(\mc{W}) \mc{W}$ is the total work averaged over many realizations of the same nonequilibrium process. Common formulations of the second law only state that $W_\mrm{irr}\geq 0$ where the equality sign is attained for quasistatic, infinitely slow processes. For \textit{finite-time} processes the irreversible work is strictly positive, and thus the natural quest for the optimal process arises, that is to identify the process that dissipates the least amount of work.

To this end, three general avenues of research were pursued during the last three decades. One approach stipulated the field of finite-time thermodynamics \cite{salamon_1982,salamon_1983,andresen_1984}, while a second one focuses on accurate estimates of free energy differences in computer simulations \cite{reinhardt_1993,dekoning_1997,zuckerman_2004,dellago_2010}. More recently the study of so-called fluctuation theorems has attracted a lot of attention. In particular, the theorems of Jarzynski \cite{jarzynski_1997,jarzynski_1997_PRE} and Crooks \cite{crooks_1998,crooks_1999} found wide-spread prominence in virtually all areas of research in classical and quantum thermodynamics \cite{bustamante_2005,campisi_2011}, as for instance, in biophysics \cite{liphardt_2002,collin_2005}, in chemical physics \cite{zimanyi_2009}, in linear response theory \cite{andrieux_2004,andrieux_2008}, and also to improve numerical algorithms \cite{sun_2004,jarzynski_2008,ballard_2012,sivak_2013}.

The present paper proposes an approach within the paradigm of finite-time thermodynamics. Imagine a thermodynamic system with Hamiltonian $H(\lambda)$, where $\lambda$ is an external control parameter. Then we ask for the optimal protocol $\lambda^*_t$ that drives the system from $H(\lambda_0)$ to $H(\lambda_\tau)$ such that the least amount of work is dissipated during finite time $\tau$. In previous works this question has been addressed within two independent approaches: If full information about the microscopic properties of the system is available the dynamics can be described by a Langevin equation \cite{seifert_2007,seifert_2008,then_2008,aurell_2011}, whereas  phenomenological treatments rely on methods of linear response theory \cite{dekoning_2005,lindberg_2009,crooks_2012,crooks_2012_PRE}. Generally, solutions obtained within the microscopic treatment are exact and valid for any kind of driving, fast and slow, strong and weak, whereas phenomenological treatments have been restricted to weak, slow driving. Nevertheless, linear response results have been more promising as only very few examples can be treated analytically in the microscopic description. In addition, descriptions by methods of linear response theory led to the discovery of new effects, as for instance geometric magnetism \cite{campisi_2012,thinga_2014}.

In the following we will derive an analytical and tractable expression for the irreversible work for slow, but not necessarily weak driving. To this end, we will show how common tools of linear response theory can be applied to \textit{slowly driven systems}. These are systems, whose driving is much slower than the relaxation induced by the thermal environment. As main results, we not only obtain an integral expression for $W_\mrm{irr}$, but also show how the optimal driving protocols $\lambda^*_t$ can be obtained from variational calculus. It will turn out that our approach significantly broadens the scope of previous treatments.

\paragraph*{Outline}

The paper is organized as follows: In Sec.~\ref{sec:slow} we motivate our approach and then derive an expression for $W_\mrm{irr}$ within a generalized linear response theory. Section~\ref{sec:corr} is dedicated to obtaining an analytical expression for the correlation time. Finally, in Sec.~\ref{sec:example} we present various examples for which the optimal protocols can be obtained analytically, before we conclude the analysis with a few remarks in Sec.~\ref{sec:con}.

\section{The regime of slowly varying processes \label{sec:slow}}

The only processes that are fully describable by means of classical thermodynamics are quasistatic processes \cite{callen_85}. Such processes, however, are only of limited relevance for practical purposes as they are infinitely slow. Moreover, they only describe situations, in which the state of a thermodynamic system evolves as a succession of \textit{equilibrium} states. All real physical processes operate in finite time and are described by a temporal succession of \textit{equilibrium} and \textit{nonequilibrium} states. 

\subsection{Physical motivation -- Biomolecule experiments}

Almost 20 years ago Jarzynski achieved a major breakthrough by relating real, finite-time processes with their quasistatic counterpart. In particular he showed \cite{jarzynski_1997} that
\begin{equation}
\label{q01}
\overline{\e{-\beta \mc{W}}}=\e{-\beta \Delta F}
\end{equation}
where $\mc{W}$ is the work performed in a single realization of a nonequilibrium process, $\beta$ is the inverse temperature, and $\Delta F$ the free energy difference. The bar denotes here an average over an ensemble of realizations weighted by the probability distribution $\mc{P}(\mc{W})$. In essence, the Jarzynski equality \eqref{q01} allows to determine the work performed during a quasistatic process, the free energy difference, from an average over an ensemble of finite-time realizations of the same process.

The Jarzynski equality \eqref{q01} was verified in a conceptually simple biomolecule experiment \cite{liphardt_2002}. The ends of an RNA molecule are attached to microscopic beads, which allow to 'pull the molecule' apart. Due to the internal structure of RNA one observes folding and unfolding behavior. To study Eq.~\eqref{q01} the following experiment is performed: The RNA molecule is brought into contact with the beads, and let to relax into its equilibrium state. Then, the molecule is pulled apart, while the applied force and the length of the molecule are recorded. The thermodynamic work $\mc{W}$ can be determined by basically evaluating 'force $\times$ displacement'. The left side of Eq.~\eqref{q01} is then simply obtained by running the same experiment many times. For the right side, however, one has to identify the quasistatic process. To this end, it is useful to notice that every reversible process coincides with a quasistatic process \cite{callen_85}. In the RNA pulling experiment \cite{liphardt_2002} a reversible process is identified if the force-displacement graph recorded during the unfolding process coincides with the graph recorded during the re-folding process. Nonequilibrium, irreversible processes show a significant hysteresis in the unfolding-folding graph \cite{liphardt_2002}.

\begin{figure}
\includegraphics[width=0.48\textwidth]{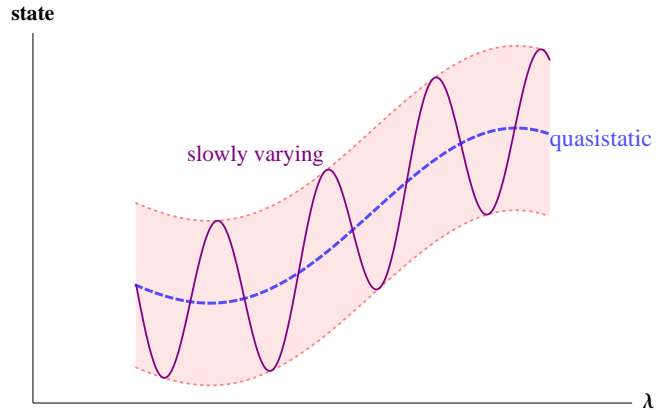}
\caption{(color online) \textbf{Schematic representation of a slowly varying process:} The purple, soild line represents a slowly varying process in the vicinity of the quasistatic process (blue, dashed line). The shaded area illustrates the \textit{linear regime}. A \textit{slowly varying process} drives the system only slightly out of equilibrium so that at all instants the system rapidly relaxes back to equilibrum; this is illustrated by the oscillations around the quasistatic path.  \label{fig00}}
\end{figure}

Nevertheless, these experiments were run in finite-time, and even during the apparently reversible process small amounts of work dissipated into the environment. In the following, our aim is to quantify these irreversible contributions. To this end, we will introduce and analyze the notion of a \textit{slowly varying process}, i.e., processes that are within \textit{the linear regime} around the quasistatic process, cf. the sketch in Fig.~\ref{fig00}.

\subsection{Irreversible work from linear response theory}

Imagine a thermodynamic system of interest that is in contact with a thermal environment. Then its equilibrium state is given by the Boltzmann-Gibbs distribution,
\begin{equation}
\label{q02}
p_\mrm{eq}(\mb{\Gamma};\,\lambda)=\e{-\beta H(\mb{\Gamma};\,\lambda)}/Z(\beta,\lambda)\,,
\end{equation}
where $Z(\beta,\lambda)$ is the partition function, $Z(\beta,\lambda) = \int \td\mb{\Gamma}\,\e{-\beta H(\mb{\Gamma};\,\lambda)}$, and $\mb{\Gamma}$ denotes a point in phase space. Note that generally $H(\mb{\Gamma};\,\lambda)$ describes the total system, which consists of system of interest and thermal reservoir. However, for the present analysis we only need that for all $\lambda$ there is a well-defined equilibrium state \eqref{q02}, where $\beta$ is the (inverse) temperature of the heat bath.

By $\lambda$ we denote an external control parameter, as for instance volume, pressure, magnetic field, etc. Work is performed by the system under study if $\lambda$ is changed according to an externally predefined protocol, $\lambda(t)$. It will prove convenient to write,
\begin{equation}
\label{q03}
\lambda(t)\equiv\lambda_0+\delta\lambda\,g(t)\,,
\end{equation}
where $g(t)$ obeys $g(0)=0$ and $g(\tau)=1$. Thus, $\lambda(t)$ is varied from $\lambda(0)=\lambda_0$ to $\lambda(\tau)=\lambda_0+\delta\lambda$ during time $\tau$. For infinitely slow variation, i.e., in the limit $\tau\rightarrow \infty$ the work performed by the system is given by the free energy difference,
\begin{equation}
\label{q04}
\Delta F \equiv F(\beta;\lambda_0+\delta\lambda) - F(\beta;\lambda_{0})
\end{equation}
where we additionally have, $F(\beta;\lambda)=-1/\beta\,\ln Z(\beta,\lambda)$. The maximum work theorem, or more fundamentally the Jarzynski equality \eqref{q01} now predicts that for all finite values of $\tau$ we have
\begin{equation}
\label{q05}
W_\mrm{irr}=W-\Delta F\geq 0\,,
\end{equation}
which means that for all realistic processes irreversible work $W_\mrm{irr}$ is dissipated into the environment. It is worth emphasizing that $W=\overline{\mc{W}}$ is an average over an ensemble of realizations of the same process. The probability for a single realization is given by, $\mc{P}(\mc{W})=\la \de{\mc{W}-\mc{W}[\mb{\Gamma}_t]}\ra$, where $\mb{\Gamma}_t$ is a trajectory in phase space \cite{chernyak_2005,deffner_2011}. This means that $\mc{P}(\mc{W}) $ can be obtained from an average over all possible paths, i.e., by a path integral average \cite{chernyak_2005,deffner_2011}. It was shown that the average thermodynamic work $W$ can also be written as \cite{jarzynski_1997_PRE}
\begin{equation}
\label{q06}
W = \int_{0}^{\tau} \td t\,\frac{\td\lambda}{\td t} \,\overline{\frac{\partial H}{\partial\lambda}}\,.
\end{equation}
In the latter equation we introduced the notation $\overline{X}$  to denote the nonequilibrium average, i.e., the average over all paths of the observable $X\equiv\pd H/\pd\lambda$. Note that Eq.~\eqref{q06} is true for any kind of driving, slow and fast, weak and strong. For the latter analysis we call $X=\partial H/\partial\lambda$ the generalized force. Note, that this definition of a \textit{generalized} force is actually minus the \textit{mechanical} force given by $-\partial H/\partial\lambda$. This choice of sign is motivated by thermodynamic considerations. The work as defined by Eq.~\eqref{q06}  is equal to the variation of the internal energy of the total system composed of system of interest plus heat bath, i.e., using Hamilton's equations, Eq.~\eqref{q06} reads \cite{deffner_jarzynski_2013}
\begin{equation}
W = \int_{0}^{\tau} \td t\,\overline{\frac{\td H}{\td t}} = \Delta U_{tot}\,.
\end{equation} 
Therefore, when $\Delta U_{tot} > 0$ the external agent has performed work and hence $W > 0$, which agrees with the sign convention we adopt in the expression (\ref{q05}) for the second law.
\begin{figure}
\includegraphics[width=.48\textwidth]{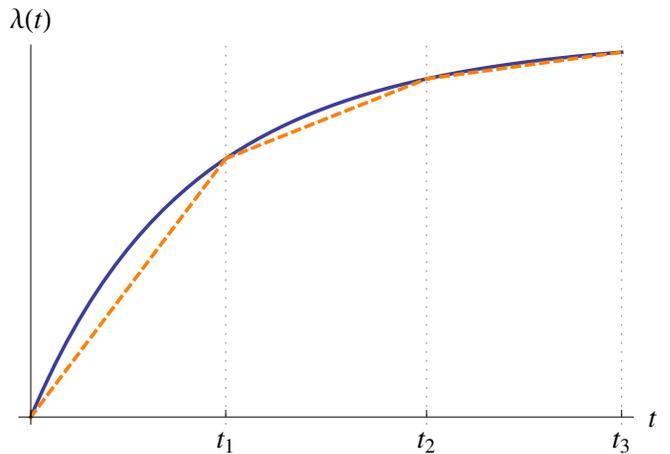}
\caption{(color online) \textbf{Typical protocol and piecewise linear approximation:} Schematic representation of a typical protocol (blue, solid line) \eqref{q03} and a piecewise linear approximation (orange, dashed line) similar to the one used to obtain the linear approximation for $W_{\mathrm{irr}}$ in Eq.~\eqref{q19}. \label{fig0}}
\end{figure}

In the remainder of this section we want to find an approximation of $W_\mrm{irr}$ for processes that are close to the corresponding  quasistatic process. Thus, mathematically we will have to find approximations, which express the state of the system being close to the equilibrium state corresponding to the instantaneous value of $\lambda(t)$.

To this end, imagine that we can separate the process of length $\tau$ into $N$ time steps of length $\delta t\equiv\tau/N$. During each of these time steps the time evolution of the protocol, described by $\lambda_n(t)=\lambda_n+\delta\lambda_n\,g_n(t)$, is then only allowed to change by $\delta\lambda_n$ for $n\in\{0,1,\dots,N\}$, where $\delta \lambda_{n}$ has to be small enough to employ methods of linear response theory for each interval. In complete analogy to the total process, $g_n(t)$ interpolates between $\lambda_n$ and $\lambda_n+\delta\lambda_n$, and therefore fulfills the boundary conditions $g_n(n\, \delta t)=0$ and $g_n((n+1)\,\delta t)=1$. 

Without loss of generality let us consider the first time interval, $0\leq t\leq\delta t$. In this case we can expand the Hamiltonian for times $t\geq 0$ in orders of $\delta\lambda_0$ and we have,
\begin{equation}
\label{q07}
H(\lambda(t)) \simeq H(\lambda_{0}) + \delta\lambda_{0}\,g_{0}(t)\,\frac{\partial H}{\partial\lambda}\bigg|_{\lambda=\lambda_{0}} + \mc{O}\left(\delta\lambda_0^2\right)\,.
\end{equation}
In the latter equation we suppressed the explicit dependence of the Hamilton on $\mb{\Gamma}$ for the sake of simplicity of notation. We further had to implicitly assume that $H(\lambda(t))$ is a regular enough function in $\lambda(t)$, so that the latter expansion is mathematically well-behaved.

It has been recently shown that dissipation originates in the lag of the dynamical state behind its corresponding equilibrium state \cite{suri_2009}. In this context 'lag' refers to the notion that nonequilibrium states generically relax into equilibrium states, if the driving is turned off. Thus, nonequilibrium states can be understood 'to lag in relaxation' behind equilibrium states. If the Hamiltonian is modulated only weakly \eqref{q07} the real nonequilibrium state lags only 'slightly' behind the instantaneous equilibrium state, and we can express the nonequilibrium average of Eq.~\eqref{q07} by means of linear response theory \cite{kubo_1957,kubo_1985,andrieux_2004,andrieux_2008},
\begin{equation}
\label{q08}
\overline{\frac{\partial H}{\partial \lambda}} =\la\frac{\partial H}{\partial \lambda}\ra_0  + \chi^{\infty}_{0}\, \delta\lambda_{0}\,g_{0}(t) - \delta\lambda_{0}\,\int_{0}^{t}\td s\,\phi_{0}(t-s)\, g_{0}(s)\,.
\end{equation}
The angular brackets, $\la X\ra_n$ denote an average of an observable $X$ over the equilibrium state for the $n$th time step,
\begin{equation}
\label{q09}
\la X\ra_{n}=\int d\mathbf{\Gamma}\,X(\mathbf{\Gamma},\lambda_{n})\exp{(-\beta H(\mathbf{\Gamma};\,\lambda_{n}))}/Z(\beta;\lambda_{n})\,.
\end{equation}
Equation~\eqref{q08} has a clear physical interpretation: The second term describes the \textit{instantaneous} response, which is due to the observable  $X=\partial H/\partial\lambda$ being a function of the external control\cite{kubo_1985}. In particular, we have
\begin{equation}
\label{q10}
\chi^{\infty}_{0} = \la \frac{\partial^{2}H}{\partial\lambda^{2}}\ra_{0}\,.
\end{equation}
The third is the so-called after-effect contribution, the \textit{delayed} response. It is governed by the \textit{response function} \cite{kubo_1985},
\begin{equation}
\label{q11}
\phi_{0}(t) = \la\{X(0),X(t)\}\ra_{0}\,,
\end{equation}
where $\{A,B\} = \partial_qA\cdot\partial_pB - \partial_pA\cdot\partial_qB$ is the Poisson bracket. Employing Kubo's formula we have $\phi_{0}(t) = -\dot{\Psi}_{0}(t)$, where $\Psi_{0}(t)$ is the \textit{relaxation function} \cite{kubo_1985}, and
\begin{equation}
\label{q12}
\Psi_{0}(t) = \beta\left( \la X(0)X(t)\ra_{0} - \la X(0)\ra^{2}_{0}\right)\,.
\end{equation}
Therefore, Eq.~\eqref{q08} can be re-written after an integration by parts as
\begin{equation}
\label{q13}
\begin{split}
\overline{\frac{\partial H}{\partial \lambda}} &=\la \frac{\partial H}{\partial \lambda}\ra_{0} 
 - \tilde{\Psi}_{0}\, \delta\lambda_{0}\,g_{0}(t) \\
 &+ \delta\lambda_{0} \int_{0}^{t}\td u\,\Psi_{0}(u)\,\frac{\td g_{0}}{\td t'}\bigg|_{t'=t-u},
\end{split}
\end{equation}
where $\tilde{\Psi}_{0} \equiv  \Psi_{0}(0) - \chi^{\infty}_{0}$.

So far we have only assumed that $\delta\lambda_0$ is small enough, so that the linear expansion of the Hamiltonian in Eq.~\eqref{q07} is a good approximation. To simplify the treatment let us further assume that $g_0(t)$ can be approximated by a linear function in $t$, which is justified for sufficiently small $\delta t$. Therefore, we have with $\td g_0/\td t\simeq \mathrm{const}$
\begin{equation}
\label{q14}
\delta\lambda_{0} \int_{0}^{t}\td u\,\Psi_{0}(u)\,\frac{\td g_{0}}{\td t'}\bigg|_{t'=t-u}\simeq\delta\lambda_{0}\,\,\frac{\td g_{0}}{\td t}\, \int_{0}^{t}\td u\,\Psi_{0}(u)\,.
\end{equation}
Furthermore, we assume that the relaxation function decays on time scales much shorter than $\delta t$. This is nothing else but an expression of the process under consideration remaining close to the quasistatic process at all times. A similar assumption is commonly employed in thermodynamics \cite{deffner_jarzynski_2013}, for any systems which is only weakly perturbed. Hence, we can write,
\begin{equation}
\label{q15}
\int^{t}_{0}\td u\,\Psi_{0}(u) \simeq \int_{0}^{\infty}\td u\,\Psi_{0}(u) \equiv \Psi_{0}(0) \,\tau^{c}_{0}\,,
\end{equation}
where $\tau^{c}_{0}$ is the {\it correlation time}, whose detailed discussion we postpone to Sec.~\ref{sec:corr}. Essentially, $ \tau^{c}_{0}$ determines the time scale over which the response vanishes, i.e., the system relaxes back to equilibrium.

Substituting Eq.~\eqref{q14} with the expression~\eqref{q15} into the integral for the work~\eqref{q06} we obtain that during the first time step the work
\begin{equation}
\label{q16}
\begin{split}
\delta W_{0} &\simeq \delta\lambda_{0} \la\frac{\partial H}{\partial \lambda}\ra_{\lambda_{0}} -\frac{(\delta\lambda_{0})^{2}}{2} \tilde{\Psi}_{0} \\
& + \delta t\,(\delta\lambda_{0})^{2} \left(\frac{\td g_{0}}{\td t}\right)^{2} \tau^{c}_{0}\, \Psi_{0}(0)
\end{split}
\end{equation}
is performed (where the integral in Eq.~\eqref{q06} was calculated assuming $\delta t$ very small). Note that for the latter equation we approximated $g_0(t)$ as a linear function, cf. Eq.~\eqref{q14}.

The task is now to identify reversible and irreversible contributions. It is easy to see that the first two terms can have either sign. In particular, reversing the arrow of time also changes the sign of the first two terms, but their absolute value remains invariant. One easily convinces oneself, that the first two terms also coincide with the free energy difference for the first time step. Therefore, we identify the first two terms in Eq.~\eqref{q16} as reversible contribution. The third term, on the other hand is always non-negative, and thus the irreversible work reads
\begin{equation}
\label{q17}
(\delta W_{0})_{\mathrm{irr}} = \delta t\,(\delta\lambda_{0})^{2} \left(\frac{\td g_{0}}{\td t}\right)^{2}\tau^{c}_{0}\, \Psi_{0}(0)\,.
\end{equation}
The latter result readily generalizes to the $n$th time step, and the general expression reads,
\begin{equation}
\label{q18}
(\delta W_{n})_{\mathrm{irr}} = \delta t\,(\delta\lambda_{n})^{2} \left(\frac{\td g_{n}}{\td t}\right)^{2}\tau^{c}_{n}\, \Psi_{n}(0)\,.
\end{equation}
It is worth emphasizing that the equilibrium state \eqref{q02} is 'updated' for each time step, and that therefore the equilibrium averages in Eq.~\eqref{q18} are taken with respect to the \textit{instantaneous} equilibrium distributions. In another words, we start each time step with an equilibrium probability distribution corresponding to a value $\lambda_{n}$. This can be understood as a consequence of the time-scale separation introduced in Eq.~\eqref{q14}. See also Nulton {\it et al.} \cite{nulton_1985} for similar assumptions.

The total irreversible work is then given by
\begin{equation}
\label{q19}
W_{\mathrm{irr}}\simeq \sum_{n=0}^{N} (\delta W_{n})_{\mathrm{irr}} = \delta t \sum_{n=0}^{N} \left(\frac{\td\lambda_{n}}{\td t}\right)^{2} \tau^{c}_{n}\,\Psi_{n}(0)\,,
\end{equation}
where $\lambda_{n}(t)$ approximates the protocol $\lambda(t)$ during the $n$th time step, see also the illustration in Fig.~\ref{fig0}. In the limit of infinitesimally small $\delta t$ we can write
\begin{equation}
\label{q20}
W_{\mathrm{irr}} = \beta\int_{0}^{\tau}\td t\,\left(\frac{\td \lambda}{\td t}\right)^{2} \tau^{c}[\lambda(t)]\,\mathcal{X}[\lambda(t)]\,,
\end{equation}
where, due to $\Psi_{\lambda}(0) = \beta ( \la X^{2}(0) \ra_{\lambda} - \la X(0) \ra_{\lambda}^{2} )$ (see Eq.~\eqref{q12}), we introduced the variance
\begin{equation}
\label{q21}
\mathcal{X}[\lambda(t)] = \left\langle \left( \frac{\partial H}{\partial\lambda}\right)^{2} \right\rangle_{\lambda(t)} - \left\langle \frac{\partial H}{\partial\lambda}\right\rangle_{\lambda(t)}^{2}\,.
\end{equation}
Equation~\eqref{q20} constitutes our first main result. The irreversible work during a process within the linear regime around a quasistatic process is determined by the correlation time, $\tau^{c}[\lambda(t)]$, and the variance, $\mathcal{X}[\lambda(t)]$, of the generalized force as properties of the instantaneous equilibrium state.

It will prove convenient to re-write the total irreversible work in analogy to the work per time step as a functional of $g(t)$ and we have
\begin{equation}
\label{q22}
W_{\mathrm{irr}} = \frac{\beta}{\tau} (\delta\lambda)^{2} \int_{0}^{1}ds\,\left(\frac{dg}{ds}\right)^{2} \tau^{c}[g(s)]\,\mathcal{X}[g(s)]\,,
\end{equation} 
which coincides with expressions derived in previous works \cite{tsao_1994,dekoning_1997,crooks_2012}. Since the functional in the previous expression {\it does not} depend on the switching time $\tau$, the optimal protocols will be independent of $\tau$, as well. Moreover, Sivak and Crooks \cite{crooks_2012} obtained an analogous expression with the 'friction tensor' being here given by $\tau^{c}[g(s)]\mathcal{X}[g(s)]$. As in their case, the optimal protocols obtained from \eqref{q22} are such that the power spent in the process is constant (see appendix \ref{ap3}).

Equation~\eqref{q22} expresses $W_{\mathrm{irr}}$ as a functional of $g(s)$ whose extrema can be found using the methods of calculus of variations \cite{gelfand}. Numerically this functional \eqref{q22} was studied previously by de Koning \cite{dekoning_2005}, where, however, the correlation time, $\tau^{c}(\lambda) $,  and the variance, $\mathcal{X}(\lambda) $, were only obtained numerically. Generally, it is rather straight forward to determine analytical expressions for $\mathcal{X}(\lambda) $ \eqref{q21}, whereas treating the correlation time is more involved. In particular, we will see in the next section that to determine $\tau^{c}(\lambda)$ knowledge about the microscopic properties of the system of interest becomes necessary.

\paragraph*{Range of validity}

As we argued earlier Eq.~\eqref{q16} implies that the work performed on the system in each time step is given by an irreversible contribution plus the free energy difference $\delta F_{n}$ between the equilibrium states for $\lambda_{n}$ and 
$\lambda_{n}+\delta\lambda_{n}$,
\begin{equation}
\delta W_{n} \simeq \delta F_{n} + (\delta W_{n})_{\mathrm{irr}}\,.
\label{qq23}
\end{equation} 
We know that for quasistatic processes the irreversible contribution has to vanish and the work is identical to $\delta F_{n}$. Therefore, we expect $(\delta W_{n})_{\mathrm{irr}}$ to be very small as the actual process deviates only slightly from the quasistatic one. It seems then natural to have the ratio $(\delta W_{n})_{\mathrm{irr}}/\delta F_{n}$ as a measure of deviations from the quasistatic limit. We investigate in the following how this limit is achieved within our approach.  Intuitively the notion of a quasistatic process implies that the time derivative of the driving function has to be very small. Equation \eqref{q13} indicates that if $\td g_{n}/\td t$ is negligible, i.e., if the process is quasistatic, the work performed by the generalized force is simply the free energy difference. Hence we have to demand not only $\delta\lambda_{n}$ but also $\td g_{n}/\td t$ to be small in order to stay close to the quasistatic limit after each time step. The question is how small $\td g_{n}/\td t$ has to be in order to fulfill these conditions.  Equation \eqref{q13}, after approximations \eqref{q14} and \eqref{q15}, can be considered as an expansion in powers of both $\delta\lambda_{0}$ and $\td g_{0}/\td t$. Then, a very simple upper bound for $\td g_{0}/\td t$ can be obtained from comparing the terms of order $\delta\lambda_{0}$ with each other when $t=\delta t$. We then obtain $\tau^{c}_{0}/\delta t\ll \gamma$ and analogously $\tau^{c}_{n}/\delta t\ll \gamma$ for the $n$th time step, where $\gamma\equiv |\tilde{\Psi}_{\lambda}/\Psi_{\lambda}(0)|$ is a constant. On the other hand, the applicability of linear response theory for each time step requires that
\begin{equation}
\frac{\lambda(t_{n}+\delta t) - \lambda(t_{n})}{\lambda(t_{n})} \simeq 
\delta t \frac{(\td \lambda/\td t)|_{t=t_{n}}}{\lambda(t_{n})} \ll 1\,,
\end{equation}
which combined with $\tau^{c}_{n}/\gamma \ll \delta t$ leads to
\begin{equation}
\left| \frac{\td \lambda/\td t}{\lambda(t)} \right| \ll \frac{\gamma}{\tau^{c}[\lambda(t)]} ,
\label{qq24}
\end{equation}
in the limit where $\delta t\to 0$. This inequality determines the class of processes for which Eq.~\eqref{q20} is valid. It mainly quantifies the time-scale separation in which Eq.~\eqref{q20} is meaningful. Early derivations invoking endoreversibility \cite{tsao_1994} and linear response \cite{dekoning_1997} did not address this point before. The same is true for the recent derivation by Sivak and Crooks\cite{crooks_2012}. Although the authors explicitly mention the range of validity of their approximations in Ref.\cite{crooks_2012}, they did not combine them to quantify how fast the system can be driven keeping Eq.~\eqref{q20} valid.  

Finally, it is worth emphasizing that a similar separation of time scales was discussed earlier in the context of finite-time thermodynamics \cite{salamon_1983}. Analogously, a \textit{generalized thermodynamic length} can be defined, which allows to 'measure' the range of validity of the linear approximation more rigorously \cite{deffner_2014_letter}.

\section{Correlation time from linear response \label{sec:corr}}

Linear response theory provides a phenomenological description surpassing the potentially involved determination of nonequilibrium states. Instead, the thermodynamic properties of a system are described by the dynamical properties of correlation functions. For all systems, that are sufficiently coupled to a thermal environment, it is plausible to assume that correlations decay rapidly. This assumption expresses our expectation that thermodynamic observables evolve independently after short transients. More mathematically this assumption is supported by considering Markovian dynamics, for which it can be shown rigorously that all correlation functions decay exponentially \cite{vankampen_2007}. Therefore, one commonly models correlation functions within linear response theory by interpolations between short time transients, the initial behavior, and an exponential decay.

\subsection{Exponential ansatz}

In the present case the crucial correlation function turns out to be an autocorrelation function \eqref{q12}, whose symmetries play an important role. To illustrate the importance of such symmetries in the phenomenological treatments, let us start with a commonly used model of simple exponential decay,
\begin{equation}
\label{q23}
\Psi_{\lambda}(t) := \Psi_{\lambda}(0)\,\e{-a |t|}\,.
\end{equation}
Inspecting Eq.~\eqref{q11}, however, it is easy to see that we have to demand that  $\lim_{t\to 0^{+}} \phi_{\lambda}(t) = 0$, since Eq.~\eqref{q12} implies that $\Psi_{\lambda}(-t) = \Psi_{\lambda}(t)$. In addition, with Kubo's formula we also have $\phi_{\lambda}(-t) = -\phi_{\lambda}(t)$. We immediately observe that the ansatz \eqref{q23} does not fulfill this property, namely $\lim_{t\to 0^{+}} \phi_{\lambda}(t) \neq 0$, and hence a more careful analysis becomes necessary. Here $\phi_{\lambda}(t)$ and $\Psi_{\lambda}(t)$ are given by Eqs.~\eqref{q11} and \eqref{q12} but with $\lambda_{0}$ replaced by a different value $\lambda$.

\subsection{Self-consistent phenomenology}

More insight can be obtained by considering the Fourier transform of the response function \cite{kubo_1972}. We have,
\begin{equation}
\label{q24}
\chi(\omega) = \chi^{\prime}(\omega) -i \chi^{\prime\prime}(\omega) \equiv \int_{0}^{\infty}\td t\,
\e{-i \omega t}\,\phi_{\lambda}(t),
\end{equation}
where $\chi^{\prime}(\omega)$ and $\chi^{\prime\prime}(\omega)$ denote the real and imaginary parts, respectively. Furthermore, due to causality the integration is chosen to start at $t=0$. The latter equation can be re-written by integration by parts to read,
\begin{equation}
\label{q25}
\int_{0}^{\infty}\td t\,\e{-i \omega t}\,\dot{\phi}_{\lambda}(t) = i \omega \chi(\omega) - \phi_{\lambda}(0)\,.
\end{equation}
Now, taking the inverse Fourier transform we obtain with $\chi^{\prime}(-\omega) = \chi^{\prime}(\omega)$ and $\chi^{\prime\prime}(-\omega) = -\chi^{\prime\prime}(\omega)$,
\begin{equation}
\label{q26}
\dot{\phi}_{\lambda}(0) = \frac{2}{\pi}\int_{0}^{\infty}\td\omega\,\left(\omega \chi^{\prime\prime}(\omega) - \phi_{\lambda}(0)\right).
\end{equation}
From the definition of the response function \eqref{q11} we conclude $\phi_{\lambda}(0) = 0$, and we also have
\begin{equation}
\label{q27}
\begin{split}
\dot{\phi}_{\lambda}(t) &= \frac{\td}{\td t}\la\{X(0),X(t)\}\ra_{\lambda} = \la\{X(0),\dot{X}(t)\}\ra_{\lambda} \\
&=\la\{X(0),\{X(t),H\}\}\ra_{\lambda}\,,
\end{split}
\end{equation}
where we used that the system evolves under the Hamiltonian $H(\mb{\Gamma};\,\lambda)$.

Comparing Eqs.~\eqref{q26} and \eqref{q27} we observe that the initial value $\dot{\phi}_{\lambda}(0)$ is determined by an equilibrium average. Therefore, $\omega \chi^{\prime\prime}(\omega)$ has to decay sufficiently rapidly to ensure convergence of the integral in \eqref{q26}. One easily convinces oneself that the exponential ansatz \eqref{q23} does not fulfill this condition, as well. The lesson to learn from this analysis is that only those phenomenological ans\"atze for $\Psi_{\lambda}(t)$  are allowed, whose short time behavior fulfills Eq.~\eqref{q26}.

Equation~\eqref{q26} together with the initial value $\phi_{\lambda}(0)=0$ belong to a hierarchy of sum rules that can be obtained by systematically integrating Eq.~\eqref{q25} \cite{kubo_1972}. In Sec.~\ref{sec:example} we will discuss various illustrative examples, and we will see that qualitative short time behavior of $\phi_{\lambda}(t)$ crucially depends on the underlying Hamiltonian. Furthermore, they provide means to self-consistently determine phenomenological expressions for $\phi_{\lambda}(t)$. For our present purposes they allow to find analytical expressions for the correlation time \eqref{q15}.

For short times the response function can be studied in terms of its Taylor expansion,
\begin{equation}
\label{q28}
\phi_{\lambda}(t) = \phi^{(0)}_{\lambda}(0) + \phi^{(1)}_{\lambda}(0) t + \phi^{(2)}_{\lambda}(0) \frac{t^{2}}{2!} +\mc{O}(t^3),
\end{equation}
where the coefficients $\phi^{(n)}_{\lambda}(0)$ are given by the equilibrium average values. We have with Eq.~\eqref{q27} 
\begin{subequations}
\label{q29}
\begin{align}
\phi^{(0)}_{\lambda}(0)& = \la \{ X(0), X(0) \} \ra_{\lambda} = 0 , \\
\phi^{(1)}_{\lambda}(0) &= \la \{ X(0), \{ X(0), H \} \} \ra_{\lambda}, \\
\phi^{(2)}_{\lambda}(0) &= \la \{ X(0), \{ \{ X(0), H \}, H\} \} \ra_{\lambda}.
\end{align}
\end{subequations}
As noted earlier, symmetries become useful. In particular, we have $\phi_{\lambda}(-t) = -\phi_{\lambda}(t)$,  and thus the coefficients $\phi^{(n)}_{\lambda}(0)$, with $n$ even, vanish. 

Now imagine that we have a certain phenomenological expression for $\phi_{\lambda}(t)$, with free parameters to be determined. Then, its short time behavior has to match Eq.~\eqref{q26} with coefficients \eqref{q29}. Generally, infinitely many parameters are necessary to capture the dynamics of $\phi_{\lambda}(t) $ for all times. For sufficiently small times, however, the short time behavior is well described, for instance, by (see appendix \ref{ap0})
\begin{equation}
\label{q30}
\Psi_{1,\lambda}(t) \equiv \Psi_{\lambda}(0)\, \e{-a_{1} t}(1+ b_{1} t)^{2},
\end{equation}
for $t>0$ (for $t < 0$ one has of course to take the absolute value of $t$) and with $a_{1}$ and $b_{1}$ being free parameters. From the latter we obtain the response function with the help of Kubo's formula, $\phi_{\lambda}(t) = -\dot{\Psi}_{\lambda}(t)$. Then, expanding  $\phi_{\lambda}(t)$ up to second order and comparing the coefficients with Eq.~\eqref{q29} we obtain
\begin{subequations}
\label{q31}
\begin{align}
\phi^{(0)}_{\lambda}(0) &= (a_{1} - 2 b_{1})\,\Psi_{\lambda}(0)=0, \\
\phi^{(1)}_{\lambda}(0) &= (-a_{1}^{2} + 4 a_{1} b_{1} -2 b_{1}^{2})\,\Psi_{\lambda}(0).
\end{align}
\end{subequations}
It is easy to see that with $\Psi_{\lambda}(0)\neq 0$ Eq.~\eqref{q31} can be solved for $ a_1$ and $b_1$ as a function of $\lambda$. Finally, an expression for the correlation time \eqref{q15} is given by 
\begin{equation}
\label{q32}
\tau^{c}(\lambda) = \frac{\alpha_{1}}{a_{1}(\lambda)} = \alpha_{1} \left( \frac{\Psi_{\lambda}(0)}{2\phi^{(1)}_{\lambda}(0)}\right)^{1/2}\,,
\end{equation}
where $\alpha_{1} = 5/2$ is numerical constant. Thus,  given any particular system  $\tau^{c}(\lambda) $ can be determined by first calculating the equilibrium averages  governing $\Psi_{\lambda}(0)$ and $\phi^{(1)}_{\lambda}(0)$ and then following the above developed 'recipe'. Other examples will be shortly presented in Sec.~\ref{sec:example} and appendix~\ref{ap2}. It is worth emphasizing that the microscopic properties of the Hamiltonian $H(\mb{\Gamma};\,\lambda)$ enter the correlation time via the equilibrium averages.

In the upper discussion we restricted ourselves to the simplest case, namely to an ansatz of only two free parameters \eqref{q30}. This ansatz approximates the dynamics of the response function sufficiently well for short enough times. By sufficiently well we mean that Eq.~\eqref{q30} fulfills Eqs.~\eqref{q29} up to second order. In  appendix~\ref{ap2} we discuss various other ans\"atze that include higher order corrections. As linear response theory is a phenomenological description, it allows for a certain 'freedom of choice'. The exponential ansatz \eqref{q23} is motivated by the study of Markovian processes. The ansatz in Eq.~\eqref{q30} is slightly more general as it, in addition, takes into account a transient short time behavior. Its monotonic decay may correspond to a situation where the system is in a critical or overdamped regime  (see appendix~\ref{ap0}). If we wanted to describe underdamped motion, we would have to include an oscillatory component, see also appendix~\ref{ap2}. In general, the choice of the phenomenological expression for $\Psi_{\lambda}(t)$ is motivated by the available information about the relaxation dynamics of the system of interest.  On the other hand, it is clear from appendix~\ref{ap2} that, apart from example $d$, different ans\"atze can lead to the same dependence of $\tau^{c}$ on $\lambda$ as long as they agree with Eq.~\eqref{q28} only upto first order. For the sake of simplicity, however, we will work with the 'simplest' ansatz, that is consistent with Eqs.~\eqref{q28} and \eqref{q29}, namely Eq.~\eqref{q30}.

Before we move on illustrating our findings with the help of analytically solvable examples, let us briefly comment on the significance of the nature of the thermal environment. Generally, the total Hamiltonian can be separated into system of interest and rest of the universe $ H(\lambda) = H_{S}(\lambda) + H_\mrm{heat}$, where the control $\lambda$ acts only on the system. It has been discussed at length in the literature that then the equilibrium averages of an arbitrary observable, $\la O\ra$, only depend on the system of interest, and the bath degrees of freedom are irrelevant in this respect \cite{zwanzig_2001}. However, we will show in appendix \ref{ap1} that the nature of heat bath does manifest itself in the expression for $\phi^{(3)}_{\lambda}(0)$. Thus, the nature of the bath enters the analysis as higher order corrections.


\section{Illustrative examples \label{sec:example}}

The remainder of this paper is dedicated to the explicit discussion of analytically solvable examples. We will start with harmonic potentials, before we generalize the analysis to anharmonic cases.  Throughout this section, we restrict ourselves to the ansatz \eqref{q30}. However, as we show in appendix~\ref{ap2}, non-exponential behavior as described by Bessel and Gaussian functions can also lead to the same dependence of the correlation time $\tau^{c}$ on $\lambda$. Therefore, the only reason to focus on \eqref{q30} is its simplicity compared to other expressions (see appendix~\ref{ap0} for its physical motivation).

\subsection{Example I: harmonic trap}

In this case the control parameter $\lambda(t)$ will either represent a time-dependent minimum or a time-dependent stiffness.

\paragraph*{Time-dependent minimum}

For a harmonic oscillator transported along a fixed direction the Hamiltonian is given by
\begin{equation}
H = \frac{p^{2}}{2m} + \frac{k}{2}(q - \lambda(t))^{2}, \label{q33}
\end{equation}
and we have $\partial H/\partial\lambda = -k(q-\lambda)$. Therefore, the variance simply reads
\begin{equation}
\mathcal{X}(\lambda) = k^{2}\left(\langle(q-\lambda)^{2}\rangle_{\lambda}-\langle(q-\lambda)\rangle_{\lambda}^{2}\right) = k/\beta. \label{q34}
\end{equation}
whereas the response function coefficient becomes $\phi^{(1)}_{\lambda}(0) = \langle\left\{X(0),\left\{X(0),H\right\}\right\} \rangle_{\lambda} = k^{2}/m$. With the phenomenological ansatz \eqref{q30} for the relaxation function we obtain for the correlation time
\begin{equation}
\tau^{c}(\lambda) = \alpha_{1}\left( \frac{\Psi_\lambda(0)}{2 \phi^{(1)}_{\lambda}(0)} \right)^{1/2}= \alpha_{1}\sqrt{\frac{m}{2k}}. \label{q35}
\end{equation}
Inspecting Eqs.~(\ref{q34}) and (\ref{q35}) we observe that neither $\mathcal{X}$ nor $\tau^{c}$ depend explicitly on $\lambda$. Moreover, the reversible part of the work vanishes as the partition function remains invariant when $\lambda$ is changed. Therefore, we obtain for the irreversible work
\begin{equation}
W_{\mathrm{irr}} = \alpha_{1}\sqrt{\frac{k m}{2}}\, \frac{(\delta\lambda)^{2}}{\tau} \int_{0}^{1}\td s\,\left(\frac{\td g}{\td s}\right)^{2}. \label{q36}
\end{equation}
It is easily shown that the extremum of the functional above simply reads $g^{*}(s) = s$. This result coincides with the results obtained previously \cite{seifert_2007,seifert_2008}, apart from initial and final steps and delta peaks. In appendix \ref{ap2} we show that any phenomenological ansatz for $\Psi_{\lambda}(t)$ compatible with Eq.~\eqref{q26} yields the same results apart from a numerical prefactor $\alpha_{1}$.

\paragraph*{Time-dependent stiffness}

As a second example we consider a harmonic oscillator with time-dependent spring constant. Hence, the Hamiltonian can be written as
\begin{equation}
H = \frac{p^{2}}{2m} + \lambda(t)\frac{q^{2}}{2}, 
\label{q37}
\end{equation}
and we have $\partial H/\partial\lambda = q^{2}/2$. In this case, the variance reads
\begin{equation}
\mathcal{X}(\lambda) =\frac{1}{4}\left(\langle q^{4}\rangle_{\lambda} - \langle q^{2}\rangle_{\lambda}^{2}\right) = \frac{1}{2}\left(\beta \lambda \right)^{-2} ,
\label{q38}
\end{equation}
and the response function coefficient becomes $\phi^{(1)}_{\lambda}(0) = \langle q^{2} \rangle_{\lambda}/m = 1/\beta m\lambda $. Then, the correlation time is 
\begin{equation}
\tau^{c}(\lambda) = (\alpha_{1}/2) \sqrt{m/\lambda}.
\label{q38.1}
\end{equation}
This can be understood intuitively: in the case of the driven harmonic oscillator the characteristic time scale is determined by the period of the harmonic motion for a given $\lambda$. In appendix \ref{ap2} we argue that we obtain qualitatively the same behavior for {color{red} different} phenomenological ansatz for the relaxation function.
\begin{figure}
\includegraphics[width=.48\textwidth]{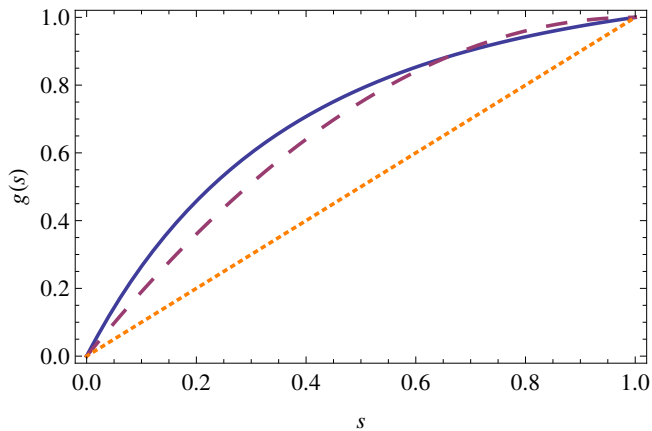}
\caption{(color online) Comparison between (\ref{q40}) (blue, solid line) for $\lambda_{0}=4.0$ and $\delta\lambda=-3.5$, a linear (orange, dotted line) and a quadratic (purple, dashed line) protocol, $g(s)=-s^{2}+2 s$.\label{fig1}}
\end{figure}
\begin{figure}
\includegraphics[width=.48\textwidth]{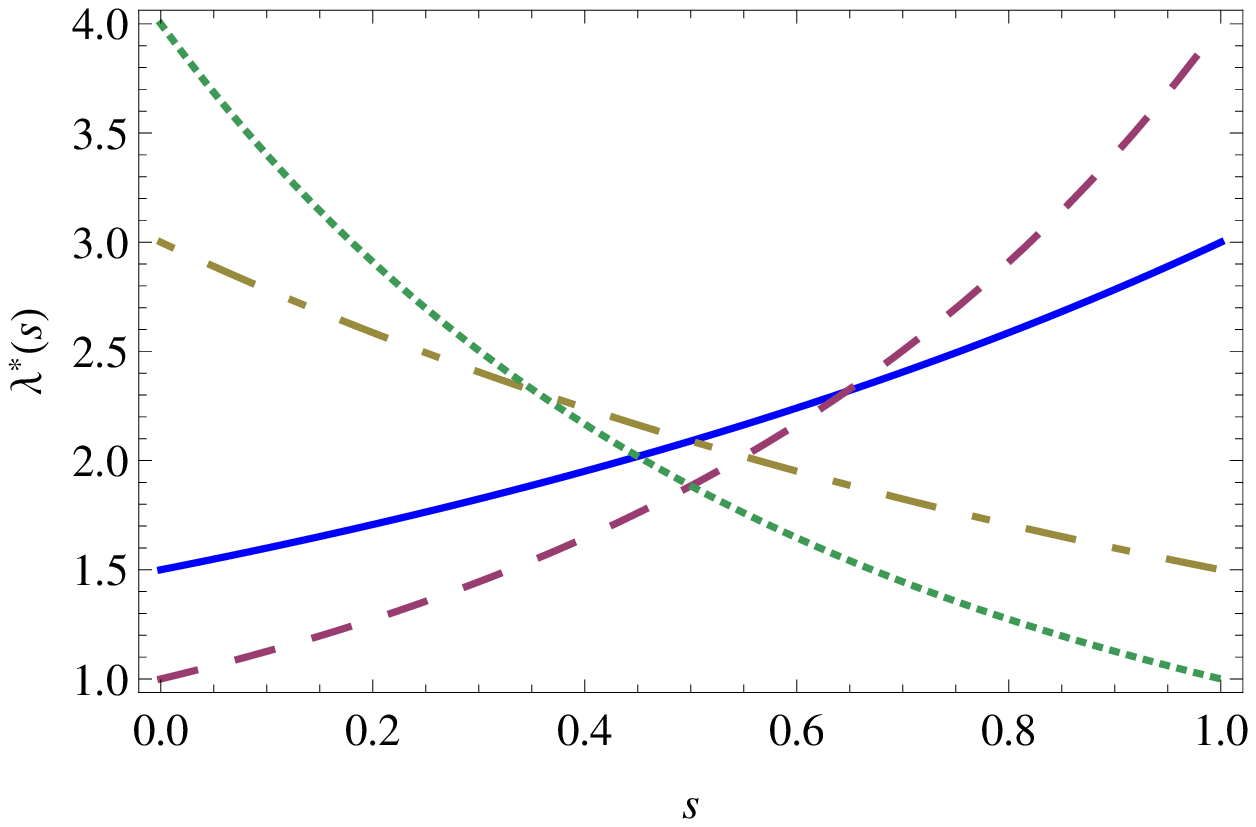}
\caption{(color online) Optimal protocols $\lambda^{*}(s)=\lambda_{0} + \delta\lambda g^{*}(s)$, with $g^{*}(s)$ given by (\ref{q40}), for different values of ($\lambda_{0}$, $\delta\lambda$): (1.5, 1.5) (blue, solid line), (1.0, 3.0) (purple, dashed line), ($3.0$, $-1.5$) (ocher, dotted-dashed line) and (4.0, $-3.0$) (green, dotted line).
\label{fig2}}
\end{figure}
The irreversible work $W_{\mathrm{irr}}$ becomes
\begin{equation}
\begin{split}
W_{\mathrm{irr}} = \frac{\alpha_{1}\,\sqrt{m}}{4\,\sqrt{\lambda_{0}}\beta\tau}
 \frac{(\delta\lambda)^2}{\lambda_{0}^2}\int_{0}^{1}\td s\,\left(\frac{\td g}{\td s}\right)^{2} \left(1 + \frac{\delta\lambda}{\lambda_{0}} g(s)\right)^{-5/2},
\label{q39}
\end{split}
\end{equation}
According to appendix \ref{ap3}, the minimum is found for the protocol
\begin{equation}
g^{*}(s) = -\frac{\lambda_{0}}{\delta\lambda} + \frac{1}{A (s+B)^{4}}, \label{q40}
\end{equation}
where $A$ and $B$ are free constants to be determined by the boundary conditions $g^{*}(0)=0$ and $g^{*}(1) = 1$. 

Choosing $\delta\lambda = -3.5$ and $\lambda_{0} = 4.0$ as used by de Koning\cite{dekoning_2005}  our analytical result \eqref{q40} qualitatively agree with the numerical outcome published earlier. As in Fig. 5 of Ref. \cite{dekoning_2005}, Figure \ref{fig1} shows the optimal protocol in terms of $g(s)$ although there it was called $\lambda(s)$ (see Eq.~(13) there). By qualitative agreement we mean that both curves increase monotonically with $s$ (although de Koning's result seems to increase faster than ours) and have the same concavity. Figure \ref{fig1} also shows a linear and a quadratic protocol that fulfill the same boundary conditions. The comparison between $W_{\mathrm{irr}}^{*}$ along $g^{*}(s)$ and $W_{\mathrm{irr}}^{\mathrm{lin}}$ and $W_{\mathrm{irr}}^{\mathrm{quad}}$ along the linear and quadratic protocols leads to $W_{\mathrm{irr}}^{*}/W_{\mathrm{irr}}^{\mathrm{lin}}\approx 0.59$ and $W_{\mathrm{irr}}^{*}/W_{\mathrm{irr}}^{\mathrm{quad}}\approx 0.91$.

Finally, Fig.~\ref{fig2} illustrates (\ref{q40}) for various values of $\lambda_{0}$ and $\delta\lambda$. Again, apart from initial and final jumps, these optimal protocols agree qualitavely with those of Schmiedl and Seifert\cite{seifert_2007} obtained in the overdamped regime (see Fig. 1(a) there).


\subsection{Example II: anharmonic trap}

We continue with the simplest anharmonic potential. For these situations earlier approaches lead to exact nonlinear integro-differential equations \cite{seifert_2007,seifert_2008}, whereas here it is still feasible to solve the  Euler-Lagrange equation analytically.

\paragraph*{Time-dependent minimum}

In complete analogy with the harmonic case we start with a transport process. Thus, we have
\begin{equation}
H = \frac{p^{2}}{2m} + \frac{k}{4}(q - \lambda(t))^{4}, \label{q41}
\end{equation}
that yields $\partial H/\partial\lambda = -k(q-\lambda)^{3}$. Accordingly, the variance reduces to
\begin{equation}
\mathcal{X}(\lambda) = 6\frac{\Gamma(3/4)}{\Gamma(1/4)} \left(\frac{k}{\beta^{3}}\right)^{1/2}, 
\label{q42}
\end{equation}
where $\Gamma$ is the Gamma function, and the response function coefficient reads
\begin{equation}
\phi^{(1)}_{\lambda}(0) = \frac{9 k}{\beta m}\,.
\label{q43}
\end{equation}
Therefore, the correlation time can be written as
\begin{equation}
\tau^{c}(\lambda) = \alpha_{1} \left(\frac{\Gamma(3/4)}{3 \Gamma(1/4)}\right)^{1/2}
m^{1/2}\left(\frac{\beta}{k}\right)^{1/4}\,.
\label{q44}
\end{equation}
In contrast to the previous examples, $\tau^{c}$ depends on the temperature, which to be expected as the system is nonlinear. Nevertheless, in complete analogy with the harmonic potential, $\mathcal{X}$ and $\tau^{c}$ do not depend on $\lambda$, and $\Delta F = 0$. Similarly, we show in the appendix \ref{ap2} that different choices of $\Psi_{\lambda}(t)$ yield the same dependence in $\beta$, $k$ and $\lambda$ as long as they fulfill Eq.~\eqref{q26}. 

Collecting terms we obtain for the irreversible work
\begin{equation}
W_{\mathrm{irr}} = \tilde{\alpha_{1}} (\delta\lambda)^{2}\, \frac{\sqrt{m}}{\tau} \left( \frac{k}{\beta} \right)^{1/4} \int_{0}^{1}\td s\,
\left( \frac{\td g}{\td s} \right)^{2},
\label{q45}
\end{equation}
where $\tilde{\alpha_{1}} = \frac{6\alpha_{1}}{\sqrt{3}} \left(\frac{\Gamma(3/4)}{\Gamma(1/4)}\right)^{3/2}$. As before the minimum is simply given by $g^{*}(s) = s$.

\paragraph*{Time-dependent stiffness}

Analogously to the previous example, we also investigate the Hamiltonian
\begin{equation}
H = \frac{p^{2}}{2m} + \lambda(t) \frac{q^{4}}{4},
\label{q46}
\end{equation}
where we have $\partial H/\partial\lambda = q^{4}/4$. Therefore, the variance becomes $\mathcal{X}(\lambda) = \left(2\beta\lambda\right)^{-2}$ and the response function coefficient reads
\begin{equation}
\phi^{(1)}(0)  = \frac{\Gamma(7/4)}{\Gamma(1/4)}\frac{1}{m}\left(\frac{4}{\beta\lambda}\right)^{3/2}\,.
\label{q47}
\end{equation}
Accordingly, the correlation time becomes
\begin{equation}
\tau^{c}(\lambda) = \frac{\alpha_{1}}{8} \left(\frac{\Gamma(1/4)}{\Gamma(7/4)}\right)^{1/2} m^{1/2} 
\left(\frac{\beta}{\lambda}\right)^{1/4}.
\label{q48}
\end{equation}
In contrast to the harmonic case $\tau^{c}$ shows two distinct features: a power law dependence on 
$\lambda$, which clearly reflects the shape of the potential and a temperature dependence. 
\begin{figure}
\includegraphics[width=.48\textwidth]{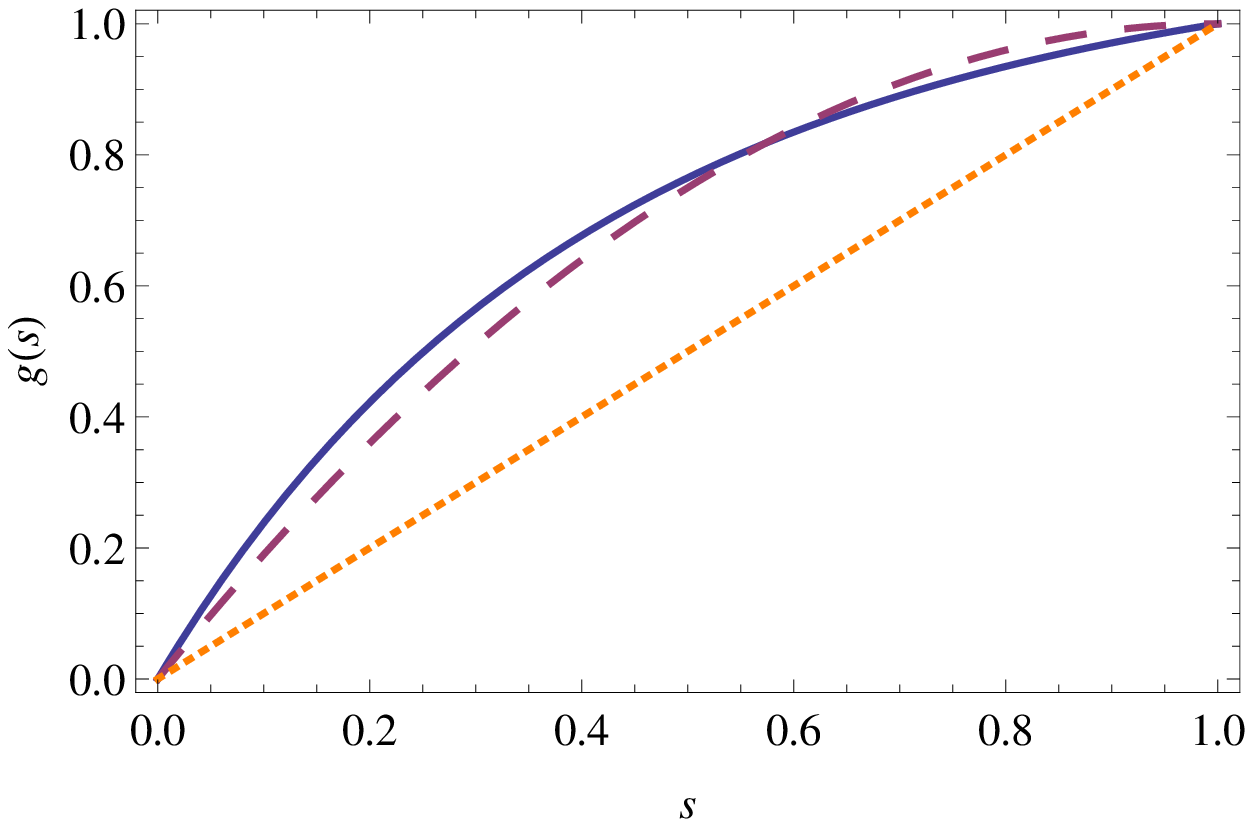}
\caption{(color online) Comparison between (\ref{q50}) (blue, solid line) for $\lambda_{0}=4.0$ and $\delta\lambda=-3.5$, a linear (organe,dotted line) and a quadratic (purple, dashed line) protocol, $g(s)=-s^{2}+2 s$.\label{fig3}}
\end{figure}
\begin{figure}
\includegraphics[width=.48\textwidth]{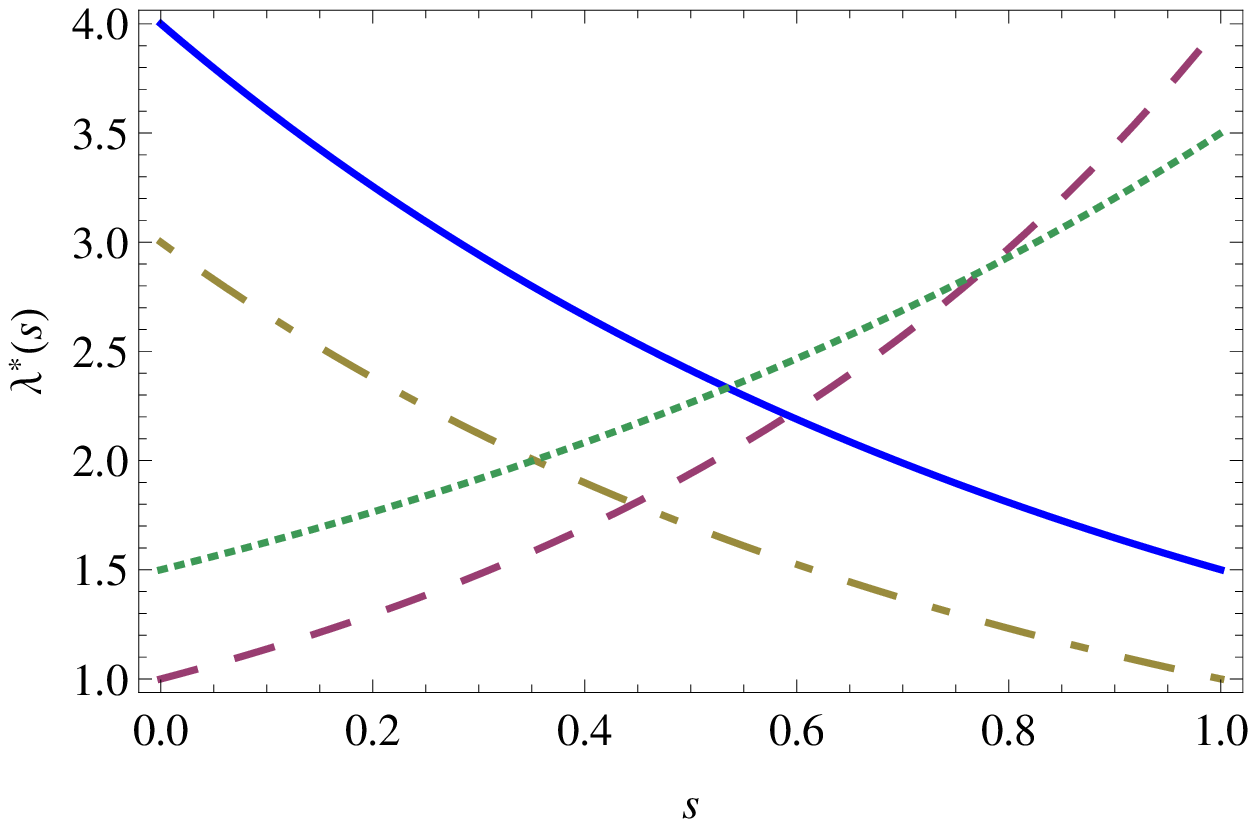}
\caption{(color online) Optimal protocols $\lambda^{*}(s)=\lambda_{o} + \delta\lambda g^{*}(s)$, with $g^{*}(s)$ given by (\ref{q50}), for different values of ($\lambda_{0}$, $\delta\lambda$): (4.0, $-2.5$) (blue, solid line), (1.0, 3.0) (purple, dashed line), (3.0, $-2.0$) (ocher, dotted-dashed line) and (1.5, 2.0) (green, dotted line).
\label{fig4}}
\end{figure}

As before, collecting expressions yields for the irreversible work
\begin{equation}
\begin{split}
W_{\mathrm{irr}} = \bar{\alpha_{1}}  \frac{(\delta\lambda)^2}{\lambda_{0}^{9/4}} \frac{m^{1/2}}{\beta^{3/4}\tau}
\int_{0}^{1}\td s\,\left( \frac{\td g}{\td s} \right)^{2} \left( 1 + \frac{\delta\lambda}{\lambda_{0}} g(s)\right)^{-9/4},
\end{split}
\label{q49}
\end{equation}
where $\bar{\alpha_{1}} = \frac{\alpha_{1}}{32}\left( \frac{\Gamma(1/4)}{\Gamma(7/4)} \right)^{1/2}$. The minimum of Eq.~(\ref{q49}) is again obtained from the Euler-Lagrange equation and reads (see appendix \ref{ap3})
\begin{equation}
g^{*}(s) = -\frac{\lambda_{0}}{\delta\lambda} + \frac{1}{A (s+B)^{8}},
\label{q50}
\end{equation}
where $A$ and $B$ are constants to be determined using the boundary conditions $g^{*}(0)=0$ and $g^{*}(1)=1$.

In Fig.~\ref{fig3} we illustrate Eq.~(\ref{q50}) for $\delta\lambda = -3.5$ and $\lambda_{0} = 4.0$. It also shows a linear and a quadratic protocol that fulfill the same boundary conditions. The comparison between $W^{*}_{\mathrm{irr}}$ along (\ref{q50}) and $W^{\mathrm{lin}}_{\mathrm{irr}}$ and $W^{\mathrm{quad}}_{\mathrm{irr}}$ along, respectively, the linear and quadratic paths furnish $W^{*}_{\mathrm{irr}}/W^{\mathrm{lin}}_{\mathrm{irr}} \approx 0.65$ and $W^{*}_{\mathrm{irr}}/W^{\mathrm{quad}}_{\mathrm{irr}} \approx 0.92$. If we compare $W^{\mathrm{harm}}_{\mathrm{irr}}$ computed from using (\ref{q40}) in (\ref{q49}) and $W^{*}_{\mathrm{irr}}$, we obtain 
$W^{*}_{\mathrm{irr}}/W^{\mathrm{harm}}_{\mathrm{irr}}\approx 0.99$. Figure \ref{fig4} shows (\ref{q50}) for different values of 
$\lambda_{0}$ and $\delta\lambda$.


\subsection{Discussion}

For the latter examples we worked with the phenomenological ansatz for the relaxation function introduced above in Eq.~\eqref{q30}. In appendices \ref{ap1} and \ref{ap2}  we show that choosing another ansatz for $\Psi_{\lambda}(t)$ seems to lead to the same qualitative results. Consequently, optimal driving for underdamped and overdamped dynamics are identical within our approximations. Therefore, a comparison with the exact results \cite{seifert_2007,seifert_2008} is not immediate. However, we do observe that our results, cf. Fig.~\ref{fig2}, are in qualitative agreement with optimal driving protocols obtained from numerical analyses \cite{dekoning_2005}. 
\begin{figure}
\includegraphics[width=0.48\textwidth]{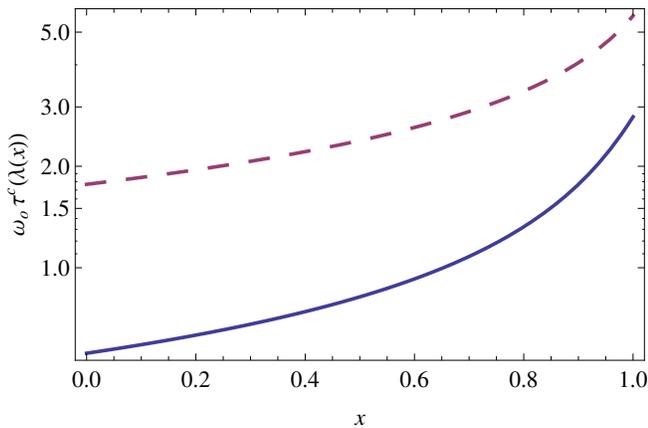}
\caption{(color online) Comparison of the correlation time for the harmonic oscillator for two response functions. $\tau_c=\alpha_1\sqrt{m/\lambda}$ for Eq.~\eqref{q30} (purple, dashed line) and  Eq.~(\ref{q52}) (blue, solid line) in logarithmic scale for decreasing $\lambda$. $\omega_{0}\equiv (\lambda_{0}/m)^{1/2}$ and $\lambda(x)\equiv \lambda_{o} + \delta\lambda\, x$ with $\lambda_{0}=10.0$ and $\delta\lambda=-9.0$.\label{fig5}}
\end{figure}

\paragraph*{Quantitative comparison with exact results}

To gain further insight a quantitative comparison of our results with the analytically exact study of Schmiedl and Seifert \cite{seifert_2007} is instructive. As a case study let us return to the harmonic trap with time-dependent stiffness \eqref{q37}. In this case the exact expression for irreversible work reads \cite{seifert_2007},
\begin{equation}
\label{quant1}
W_\mrm{exact}=\frac{1}{2}\int_0^\tau\td t\, \dot{\lambda}(t)\,w(t)+\frac{1}{2\beta}\,\lo{\frac{\lambda_0}{\lambda_0+\delta\lambda}}\,.
\end{equation} 
In the latter equation $w(t)$ denotes the mean square displacement, $w(t)=\la q^2(t)\ra$. For the remainder of this paragraph we will work in units where $\beta=1$. It has been shown by Schmiedl and Seifert \cite{seifert_2007} that for optimal driving we have 
\begin{equation}
\label{quant2}
w^*(t)=\left(1+ c\, t\right)^2/\lambda_0\,,
\end{equation}
where $c$ is a constant that depends on the initial stiffness, $\lambda_0$, the variation, $\delta\lambda$, and the switching time $\tau$,
\begin{equation}
\label{quant3}
c=\frac{1}{\tau}\,\frac{-1-\tau\,(\lambda_0+\delta\lambda)+\sqrt{1+2\lambda_0\,\tau+\lambda_0(\lambda_0+\delta\lambda)\tau^2}}{2+\tau\,(\lambda_0+\delta\lambda)}\,.
\end{equation}
Accordingly, the exact optimal protocol is given by,
\begin{equation}
\label{quant4}
\lambda^*_\mrm{exact}(t)=\left\{
\begin{aligned}
&\lambda_0 &\forall\, t\leq 0\\
&\frac{\lambda_0-c\,(1+c\,t)}{(1+c\,t)^2} &\quad\forall\, 0<t<\tau \\
&\lambda_0+\delta\lambda &\forall\, t\geq \tau
\end{aligned}
\right.\,.
\end{equation}
The purpose of this quantitative comparison is now two-fold. On the one hand, we will compare the exact protocol \eqref{quant4} with our result from linear response theory \eqref{q40}.
\begin{figure}
\includegraphics[width=.48\textwidth]{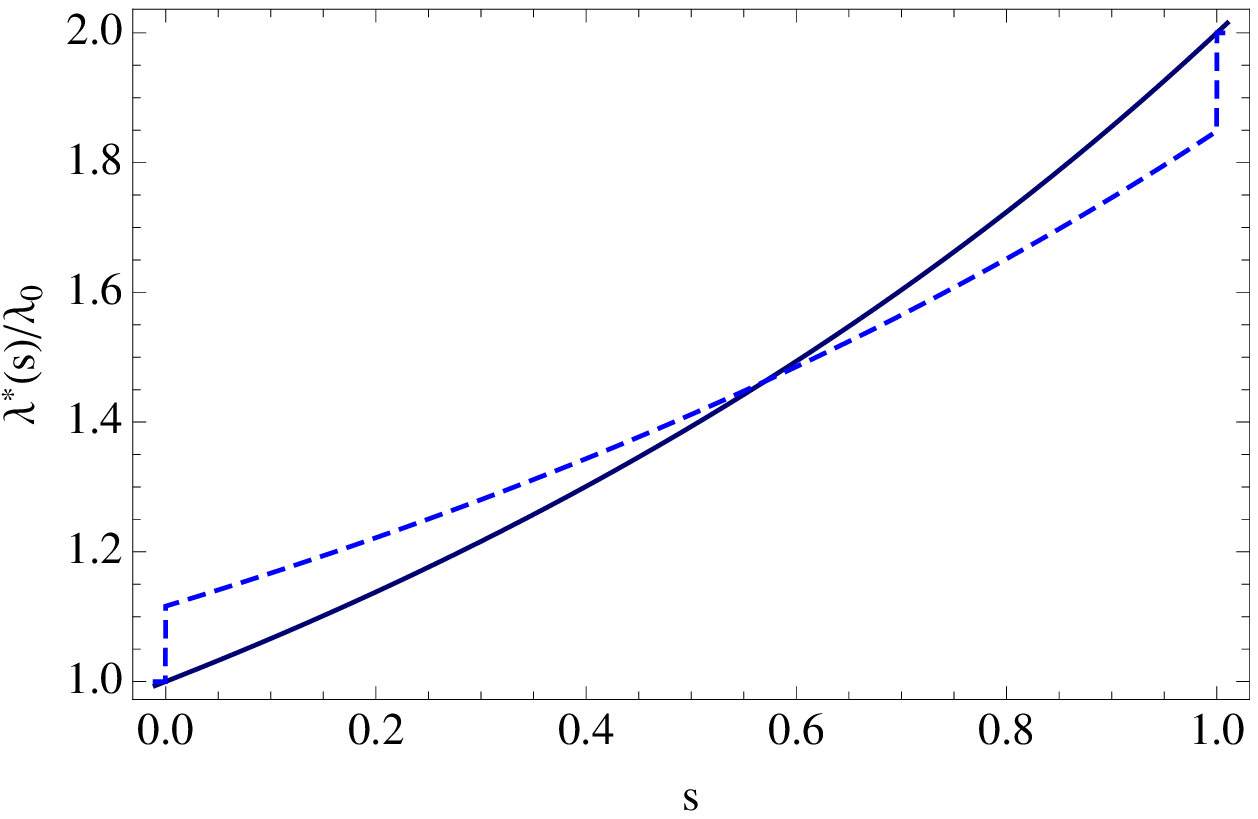}
\caption{\label{fig:compare_fast} (color online) Exact optimal protocol \eqref{quant4} (dashed line) together with the linear response result \eqref{q40} (solid line) illustrating a fast process, $\lambda_0\tau=2$; parameters are set to $\lambda_0/(\lambda_0+\delta\lambda)=2$.}
\end{figure}
\begin{figure}
\includegraphics[width=.48\textwidth]{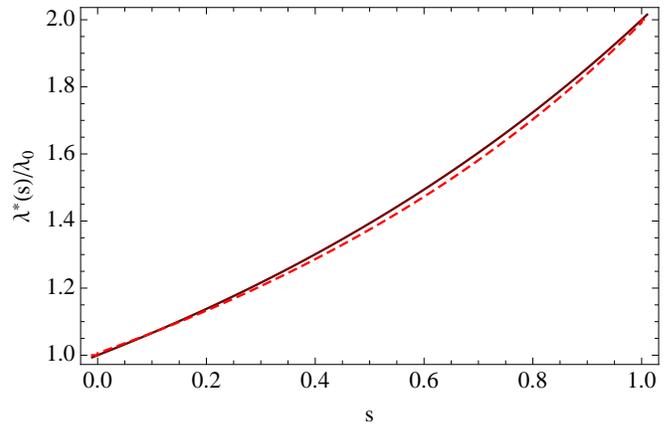}
\caption{\label{fig:compare_slow} (color online) Exact optimal protocol \eqref{quant4} (dashed line) together with the linear response result \eqref{q40} (solid line) illustrating a slow process, $\lambda_0\tau=50$; parameters are set to $\lambda_0/(\lambda_0+\delta\lambda)=2$.}
\end{figure}
On the other hand, we will check how well our protocols perform in the general case.

In Figs.~\ref{fig:compare_fast} and \ref{fig:compare_slow} we plot the exact protocol \eqref{quant4} together with our result \eqref{q40} for a fast and a slow process as quantified by the magnitude of $\lambda_0\tau$. We observe that for the slow process exact and approximate results are in very good agreement. For the fast process the exact result shows the characteristic jump behavior, which is beyond the scope of any linear response theory.

\begin{figure}
\includegraphics[width=.49\textwidth]{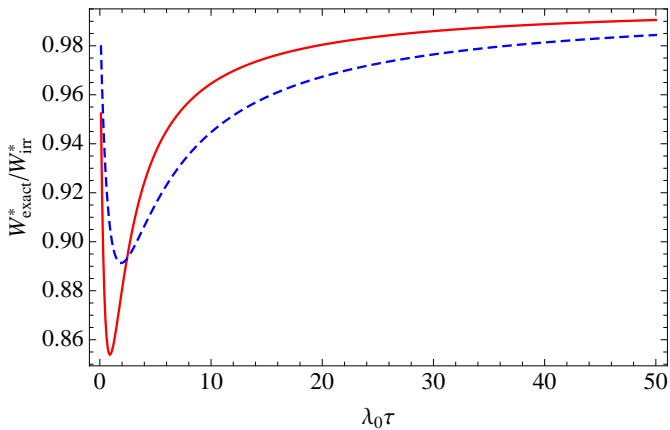}
\caption{\label{fig:ratio}(color online) Ratio of the resulting irreversible work \eqref{quant1} computed for the exact protocol \eqref{quant4}, $W^*_\mrm{exact}$, and from linear response \eqref{q40}, $W^*_\mrm{irr}$ as a function of the slowness parameter $\lambda_0\tau$ for $\lambda_0/(\lambda_0+\delta\lambda)=2$ (red, solid line) and  $\lambda_0/(\lambda_0+\delta\lambda)=0.5$ (blue, dashed line).}
\end{figure}
In order to check how well the linear response results perform in the general case we computed the exact irreversible work \eqref{quant1} for exact and approximate protocols. In Fig.~\ref{fig:ratio} we plot the ratio of the resulting values as a function of the 'slowness' parameter $\lambda_0\tau$. We observe that for slow processes, $\lambda_0\tau\gg 1$, linear response and exact results are in very good agreement, as expect from Figs.~\ref{fig:compare_fast} and \ref{fig:compare_slow}. Deviations are observed for fast processes, which cannot be described as 'slowly varying processes'.

\paragraph*{Higher order corrections}

Now, let us briefly discuss the effect of higher order corrections in the correlation time. If $\Psi_{\lambda}(t)$ is demanded to fulfill Eqs.~\eqref{q26}-\eqref{q29} up to third order the nature of the heat bath becomes important. To this end, we analyze a few examples in appendix \ref{ap2}. In particular, if we allow for the underdamped behavior (see also Eq.~\eqref{a15}),
\begin{equation}
\label{q51}
\Psi_{5,\lambda}(t)=\Psi_{\lambda}(0)\,\e{-a_{5} t^{2}}\, \cos{(b_{5} t)}\,,
\end{equation}
we obtain for the harmonic oscillator with time-dependent frequency \eqref{q37}
\begin{equation}
\label{q52}
\tau^{c}(\lambda) = \sqrt{\frac{\pi m}{2\lambda\,(1-f(\lambda))}} \,
\e{ -\frac{f(\lambda)}{2(1-f(\lambda))}},
\end{equation}
with $f(\lambda) \equiv \sqrt{2 - \eta\omega_{D}/\lambda}/2$. In Fig.~\ref{fig5} we plot Eq.~\eqref{q52} together with the simple result obtained earlier, $\tau_c=(\alpha_1/2)\sqrt{m/\lambda}$. We observe that accounting for the bath degrees of freedom yields a slightly stronger dependence of the correlation time on the control. In another words, Figure \ref{fig5} shows that the correlation times \eqref{q38.1} and \eqref{q37} (and their derivatives) both grow in a similar way as $\lambda$ decreases. 

Finally, we point out that the heuristic arguments used to derive Eq.~\eqref{qq24} need to be discussed more carefully for the examples where $\lambda(t)$ describes a time-dependent minimum. The reason is simply that the partition function $Z(\beta,\lambda)$, and therefore the free energy, does not depend on $\lambda$. In this cases the ratio $(\delta W_{n})_{\mathrm{irr}}/\delta F_{n}$ is meaningless since $\delta F_{n} = 0$. However $\td g/\td t$ still controls the amount of irreversible work  performed along the process. We consider then for these cases the inequality
\begin{equation}
W_{\mathrm{irr}}/\langle H \rangle_{0} \ll 1,
\label{qq53}
\end{equation}
where $\langle H \rangle_{0}$ is the initial internal energy, as a criteria for staying near the quasistatic regime.
We saw in section \ref{sec:example} that the optimal protocols for time-dependent minima are linear functions. Therefore, $\td g/\td t$ is simply given by the inverse of the switching time $\tau$. Using \eqref{q36} and $g^{*}(s) = s$ in \eqref{qq53} we obtain
\begin{equation}
\tau \gg \alpha_{1} \beta (\delta\lambda)^{2} \sqrt{k m/2},
\label{qq54}
\end{equation}
for the harmonic trap. Analogously, using \eqref{q45} and $g^{*}(s)=s$ in \eqref{qq53} leads to
\begin{equation}
\tau \gg \frac{4}{3}\tilde{\alpha}_{1} \beta (\delta\lambda)^{2} \sqrt{m} \left(\frac{k}{\beta}\right)^{1/4},
\label{qq55} 
\end{equation}
for the anharmonic trap. 

\section{Concluding remarks \label{sec:con}}

In the present analysis we used methods of linear response theory to describe slowly varying processes, i.e., processes that operate in the linear regime around the quasistatic process. This allowed us to derive a mathematically simple functional for the irreversible work, from which optimal processes can be identified.

It turns out that the irreversible work is governed by the correlation time and the fluctuations of the generalized force conjugated to the control parameter. In contrast to previous work we were also able to derive analytical, closed form expressions for the correlation time. To this end, we developed a self-consistent phenomenology to obtain the relaxation function. It is worth emphasizing that our novel approach allows to determine analytical expressions for the correlation time of nonlinear systems, where the description in terms of Langevin or Fokker-Planck equations is very limited.

As illustrative examples we further studied harmonic and anharmonic oscillators. For these we found that the optimal control, i.e., the control that minimizes the irreversible work, are in qualitative agreement with results from the literature. The optimal protocols turn out to be independent of the total switching time and the temperature. Nevertheless, it still poses an open problem to reconcile the 'jump' processes reported for systems described by Langevin dynamics \cite{seifert_2007,seifert_2008}, and the completely continuous protocols from our linear response theory.

\acknowledgments
It is a pleasure to thank M. de Koning for valuable discussions and suggestions and C. Jarzynski for the hospitality during M.B.'s visit to the University of Maryland, College Park. S.D. acknowledges financial support from the National Science Foundation (USA) under grant DMR-1206971 and M.B. support from the Brazilian research agency FAPESP under the contract 2012/07429-0.

\appendix
\section{Heat bath influence}
\label{ap1}

In this appendix we have a closer look at the importance of the nature of the heat bath in our analysis. As before we consider the generalized force, $X \equiv \partial H/\partial\lambda$, which is only a function of the particle coordinate $q$, i.e., $X = X(q)$. Therefore, its Poisson bracket with any other observable $O$ reads, 
\begin{equation}
\begin{split}
\label{b1}
\{ X, O\}& \equiv  \frac{\partial X}{\partial q}\frac{\partial O}{\partial p} - \frac{\partial X}{\partial q}\frac{\partial O}{\partial p} \\
&+\sum_{k=1}^{N}\left( \frac{\partial X}{\partial q_{k}}\frac{\partial O}{\partial p_{k}} - \frac{\partial X}{\partial q_{k}}\frac{\partial O}{\partial p_{k}} \right)\\
&=\frac{\partial X}{\partial q} \frac{\partial O}{\partial p}
\end{split}
\end{equation}
where we denote here by $(q,p)$ and by $(q_{k},p_{k})$ the phase space coordinates of the system of interest and heat bath, respectively. Let us now define
\begin{equation}
\label{b2}
B^{(1)} \equiv \{X,H\} = \frac{p}{m} \frac{\partial X}{\partial q} ,
\end{equation}
where $H=H_S+H_\mrm{heat}$ the total Hamiltonian, consisting of system of interest, $H_S$, and thermal reservoir, $H_\mrm{heat}$. We also have
\begin{equation}
\label{b3}
\{ X, B^{(1)} \} = \frac{1}{m} \left( \frac{\partial X}{\partial q} \right)^{2},
\end{equation}
and we can write with Eq.~\eqref{q29},
\begin{equation}
\label{b4}
\phi^{(1)}_{\lambda}(0) = \langle \{ X, B^{(1)} \} \rangle_{\lambda} = \frac{1}{m} \la \left( \frac{\partial X}{\partial q} \right)^{2} \ra_{\lambda}.
\end{equation}
Now let us assume that the thermal reservoir can be written as an ensemble of harmonic oscillators,
\begin{equation}
\label{b5}
H_\mrm{heat} = \sum_{k=1}^{N} \left[ \frac{p_{k}^{2}}{2 m_{k}} + \frac{m_{k}\omega_{k}^{2}}{2}(q_{k} - q)^{2} \right]\,,
\end{equation}
then we can define
\begin{equation}
\label{b6}
\begin{split}
 B^{(2)} &\equiv  \{ B^{(1)}, H \}  \\
&= \frac{p^{2}}{m^{2}} \frac{\partial^{2}X}{\partial q^{2}} - \frac{1}{m}\frac{\partial X}{\partial q}
\left[ \frac{\partial H_{S}}{\partial q} - \sum_{k=1}^{N} m_{k} \omega_{k}^{2} (q_{k} - q) \right].
\end{split}
\end{equation}
Therefore, we also have
\begin{equation}
\label{b7}
\{ X, B^{(2)} \} = \frac{2 p}{m^{2}} \frac{\partial X}{\partial q} \frac{\partial^{2} X}{\partial q^{2}},
\end{equation}
which, together with Eq.~\eqref{q29}, leads to
\begin{equation}
\label{b8}
\phi^{(2)}_{\lambda}(0) = \langle \{ X, B^{(2)} \} \rangle_{\lambda} = \frac{2 \la p \ra_{\lambda}}{m^{2}} 
\la\frac{\partial X}{\partial q} \frac{\partial^{2} X}{\partial q^{2}} \ra_{\lambda} = 0,
\end{equation}
since there is no coupling between $p$ and $q$ in $H$ and $\la p \ra_{\lambda} = 0$. 

In the remainder of this appendix we will now show that the nature of the heat bath comes in third order in our treatment. To this end, let us further define
\begin{equation}
\label{b9}
\begin{split}
B^{(3)} &\equiv \{ B^{(2)}, H \}  \\
& =  \frac{\partial B^{(2)}}{\partial q} \frac{\partial H}{\partial p} - \frac{\partial B^{(2)}}{\partial p} \frac{\partial H}{\partial q} 
+ \sum_{k=1}^{N} \frac{\partial B^{(2)}}{\partial q_{k}} \frac{\partial H}{\partial p_{k}}.
\end{split}
\end{equation}
The latter can be rearranged to read
\begin{eqnarray}
\label{b10}
B^{(3)} &=& \frac{p^{3}}{m^{3}} \frac{\partial^{3} X}{\partial q^{3}} + \frac{1}{m} \frac{\partial X}{\partial q} \sum_{k=1}^{N} \omega_{k}^{2} p_{k} \nonumber \\
&-& \frac{3 p}{m^{2}} \frac{\partial^{2} X}{\partial q^{2}} \left[ \frac{\partial H_{S}}{\partial q} - \sum_{k=1}^{N} m_{k}\omega_{k}^{2} (q_{k} - q) \right] \nonumber \\
&-& \frac{p}{m^{2}} \frac{\partial X}{\partial q} \left[ \frac{\partial^{2} H_{S}}{\partial q^{2}} + \sum_{k=1}^{N} m_{k}\omega_{k}^{2} \right].
\end{eqnarray}
Thus we finally obtain
\begin{eqnarray}
\label{b11}
 \{ X , B^{(3)} \} &=& \frac{3 p^{2}}{m^{3}}\frac{\partial X}{\partial q} \frac{\partial^{3} X}{\partial q^{3}} \nonumber \\
 &-&\frac{3}{m^{2}} \frac{\partial X}{\partial q} \frac{\partial^{2} X}{\partial q^{2}} \left[ \frac{\partial H_{S}}{\partial q} - \sum_{k=1}^{N} m_{k}\omega_{k}^{2} (q_{k} - q) \right] \nonumber \\
 &-& \frac{1}{m^{2}} \left( \frac{\partial X}{\partial q} \right)^{2} \left[ \frac{\partial^{2} H_{S}}{\partial q^{2}} + \sum_{k=1}^{N} m_{k}\omega_{k}^{2} \right].
\end{eqnarray}

The microscopic parameters of the Hamiltonian \eqref{b5} can be expressed in terms of the spectral density $J(\omega)$ in the following way \cite{zwanzig_2001}
\begin{equation}
\sum_{k=1}^{N} m_{k}\omega_{k}^{2} = \frac{2}{\pi}\int_{0}^{\infty} d\omega\,\frac{J(\omega)}{\omega}.
\label{bb1}
\end{equation}
In the Ohmic regime, $J(\omega)$ is given by 
\begin{equation}
J(\omega) = \eta\omega \frac{\omega_{D}^{2}}{\omega^{2} + \omega_{D}^{2}},
\label{bb2}
\end{equation}
where $\eta$ is the friction constant and $\omega_{D}$ is a cutoff frequency. It can be shown \cite{zwanzig_2001} that this expression for $J(\omega)$ leads to an effective equation of motion for $q$ with a friction term $\eta \dot{q}$ in the limit $\omega_{D}\to\infty$.

In the case of the harmonic oscillator with time-dependent frequency \eqref{q37} we hence can write
\begin{eqnarray}
\label{b12}
\lefteqn{\phi^{(3)}_{\lambda}(0) = \langle \{ X, B^{(3)} \} \rangle_{\lambda} } \nonumber \\
&=& -\frac{4}{m^{2}} \langle q^{2}\rangle_{\lambda} \left( \lambda + \sum_{k=1}^{N} m_{k}\omega_{k}^{2} \right)
+ \frac{3}{m^{2}} \sum_{k=1}^{N} m_{k} \omega_{k}^{2} \langle q_{k} q \rangle_{\lambda} \nonumber \\
&=& -\frac{4}{m^{2}\beta}  \left( 1 + \sum_{k=1}^{N} \frac{m_{k}\omega_{k}^{2}}{4\lambda} \right) \nonumber \\
&=& -\frac{4}{m^{2}\beta}  \left( 1 +  \frac{\eta\omega_{D}}{4\lambda} \right),
\end{eqnarray}
where, from the second to the third line, we used 
\begin{equation}
\label{b13}
\langle q^{2} \rangle_{\lambda} = 1/\beta\lambda, \quad\mrm{and}\quad \langle q_{k} q \rangle_{\lambda} = 1/\beta\lambda\,,
\end{equation}
and from the third to the fourth line, we used (\ref{bb1}) and (\ref{bb2}).
The parameter $\eta\omega_{D}/\lambda$ determines the regimes of weak ($\eta\omega_{D}/\lambda\ll 1$) and strong
($\eta\omega_{D}/\lambda\gg 1$) coupling.

\section{Relaxation function from Brownian motion}
\label{ap0}

In this appendix we show an example where a very simple relaxation function can be obtained exactly. Let us consider the following Langevin equation
\begin{equation}
\ddot{q}(t) + 2\eta\,\dot{q}(t) + \omega_{o}^{2}\,q(t) = f(t)/m\,,
\label{n0}
\end{equation}
describing the motion of a particle with mass $m$ in the presence of a harmonic potential whose characteristic frequency is $\omega_{o}$. The friction constant is $\eta$ and $f(t)$ is the usual noise with mean value 
\begin{equation}
\overline{f(t)} = 0\,,
\label{n1}
\end{equation}
and correlation function given by
\begin{equation}
\overline{f(t)f(t')} = 4m\eta k_{B}T\,\delta(t-t')\,,
\label{n2}
\end{equation}
where $k_{B}$ is Boltzmann constant and $T$ is the temperature of the heat bath.

To simplify the analysis, we restrict ourselves to the situation of {\it critical} damping where $\eta = \omega_{o}$. In this case, the solution of Eq.~\eqref{n1} reads
\begin{eqnarray}
q(t) &=& q(0)\,(1+\eta t)\,\e{-\eta t} + \dot{q}(0) t\, \e{-\eta t} \nonumber \\
&+& \int_{0}^{t} \td t'\,\e{-\eta(t-t')} (t-t') \frac{f(t')}{m}\,,
\label{n3} 
\end{eqnarray}
where $q(0)$ and $\dot{q}(0)$ are the position and velocity of the particle at $t=0$. From Eqs.~\eqref{n1}, \eqref{n2} and \eqref{n3}, it is straightforward to obtain, for $t > 0$,
\begin{eqnarray}
\overline{q^{2}(t)} &=& q^{2}(0) (1+\eta t)\, \e{-2\eta t} + \dot{q}^{2}(0)\,t^{2}\,\e{-2\eta t} \nonumber\\
&+& 2 q(0) \dot{q}(0) (1+\eta t) t\,\e{-2\eta t} \nonumber \\
&+& \frac{4\eta k_{B}T}{m}\int_{0}^{t} \td t'\,\e{-2\eta(t-t')}\,(t-t')^{2}\,,
\end{eqnarray}
where, as before, the overline denotes an average over different noise realizations \cite{vankampen_2007}. Thus, the correlation function of $q^{2}(t)$ reads
\begin{equation}
\left\langle q^{2}(0)\,\overline{q^{2}(t)} \right\rangle - \langle q^{2}(0) \rangle^{2} = var\left( q^{2} \right)\,\e{-2\eta t}(1+\eta t)^{2}\,,
\label{n4}
\end{equation}
where $\langle\cdot\rangle$ denotes an average over initial conditions using a canonical distribution and $var(q^{2}) \equiv \langle q^{4}(0)\rangle - \langle q^{2}(0)\rangle^{2}$. Equation \eqref{q12} then tells us that the relaxation function $\Psi(t)$ with $X = q^{2}/2$ would be exactly proportional to \eqref{n4}. 

Motivated by the simplicity of \eqref{n4}, we show in section \ref{sec:corr} that although not exact in general the ansatz \eqref{q30} describes approximately well the relaxation function in some situations of interest. 

\section{Phenomenological expressions for the relaxation function}
\label{ap2}

This appendix is dedicated to the study of various phenomenological ans\"atze for $\Psi_{\lambda}(t)$. In particular we will see that $\tau^{c}(\lambda)$ does not change qualitatively, if $\Psi_{\lambda}(t)$ is to fulfill Eqs.~\eqref{q26}-\eqref{q29} up to first order. In another words, the details of the relaxation dynamics are irrelevant if (\ref{q29}) is the only sum rule (apart from $\phi_{\lambda}(0)=0$, of course) that has to be satisfied. In this regard, from all the expressions we present in the following, only the final one, $\Psi_{5,\lambda}(t)$, really yields different results. It is also important here to recall that not only $\phi_{\lambda}(0)=0$, but also $\phi^{(n)}_{\lambda}(0) = 0$ if $n$ is even (see comment after (\ref{q29})).

\paragraph{Bessel functions}

Let us start with an ansatz in terms $J_0(x)$, the Bessel function of first kind.
\begin{equation}
\label{a1}
\Psi_{2,\lambda}(t) \equiv \Psi_{\lambda}(0) J_{0}(a_{2} t),
\end{equation}
This expression may describe a nonexponential relaxation in an underdamped regime. From (\ref{a1}), we obtain
\begin{equation}
\label{a2}
\begin{split}
\phi_{2,\lambda}(t)& = -\frac{\td}{\td t} \Psi_{2,\lambda}(t) = \Psi_{\lambda}(0)\,\left[ \frac{a_{2}^{2} t}{2} + \mc{O}(t^3) \right]  \\
&= \phi_{\lambda}^{(1)}(0)\, t + \mc{O}(t^3).
\end{split}
\end{equation}
Thus, the correlation time can be written with $a_{2} = \sqrt{2 \phi_{\lambda}^{(1)}(0)/\Psi_{\lambda}(0)} $ as
\begin{equation}
\label{a3}
\tau^{c}_{2} = \frac{1}{a_{2}} = \left( \frac{\Psi_{\lambda}(0)}{2 \phi_{\lambda}^{(1)}(0)} \right)^{1/2}.
\end{equation}
\paragraph{Oscillatory behavior I}

Let us now consider exponential relaxation in an underdamped regime, which is phenomenologically described by
\begin{equation}
\label{a4}
\Psi_{3,\lambda}(t) \equiv \Psi_{\lambda}(0)\,\e{-a_{3} t}\, \left[ \cos{(b_{3} t)} + \frac{a_{3}}{b_{3}} \sin{(b_{3} t)} \right]\,,
\end{equation}
for $t > 0$. Therefore, we have
\begin{equation}
\label{a5}
\begin{split}
\phi_{3,\lambda}(t) &= -\frac{\td}{\td t} \Psi_{3,\lambda}(t) \\
&= \Psi_{\lambda}(0) \left[ (a_{3}^{2} + b_{3}^{2}) t - a_{3}(a_{3}^{2}+b_{3}^{2}) t^{2} + \mc{O}(t^3) \right]  \\
&= \phi_{\lambda}^{(1)}(0) \,t + \mc{O}(t^3),
\end{split}
\end{equation}
which leads to the following system of equations
\begin{subequations}
\label{a7}
\begin{align}
\phi_{\lambda}^{(1)}(0) &= \Psi_{\lambda}(0) (a_{3}^{2} + b_{3}^{2}), \label{a7a}\\
\phi_{\lambda}^{(2)}(0) &= 0 = 2 \Psi_{\lambda}(0) a_{3} (a_{3}^{2} + b_{3}^{2}). \label{a7b}
\end{align}
\end{subequations}
In general, we have $\phi_{\lambda}^{(1)}(0) \neq 0$, and Eqs.~(\ref{a7}) imply that $a_{3} = 0$. However, this is not admissible since no relaxation would occur. Therefore, we conclude that (\ref{a4}) is a good description only up to first order in the expansion (\ref{a5}). This means we will ignore (\ref{a7b}) and consider only (\ref{a7a}). Thus, the relation between $a_{3}$ and $b_{3}$ has to be introduced by hand from the knowledge about the relaxation dynamics of the system under study. For instance, if we take $b_{3} = 2 a_{3}$, we obtain $a_{2} = \sqrt{\phi_{\lambda}^{(1)}(0)/5 \Psi_{\lambda}(0)} $, and the corresponding correlation time becomes
\begin{equation}
\label{a8}
\tau^{c}_{3} = \frac{2}{5 a_{3}} = \frac{2}{5}\left( \frac{\Psi_{\lambda}(0)}{5 \phi_{\lambda}^{(1)}(0)} \right)^{1/2}.
\end{equation}

\paragraph{Gaussian response}

Now, we turn to overdamped Gaussian relaxation
\begin{equation}
\label{a10}
\Psi_{4,\lambda}(t) \equiv \Psi_{\lambda}(0)\,\e{-a_{4} t^{2}},
\end{equation}
and we have
\begin{equation}
\label{a11}
\begin{split}
\phi_{4,\lambda}(t) &= -\frac{\td}{\td t} \Psi_{4,\lambda}(t) 
= \Psi_{\lambda}(0) \left[ 2 a_{4} t - 2 a_{4}^{2} t^{3} + \mc{O}(t^5) \right] \\
&= \phi_{\lambda}^{(1)}(0) t + \phi_{\lambda}^{(3)}(0)\frac{t^{3}}{3!} + \mc{O}(t^5)\,.
\end{split}
\end{equation}
We immediately observe that while the ansatz works up to second order we cannot match the first and third order coefficients simultaneously. Therefore we conclude $a_{4} =  \phi_{\lambda}^{(1)}(0)/2 \Psi_{\lambda}(0) $, and the corresponding correlation time becomes
\begin{equation}
\label{a12}
\tau^{c}_{4} = \frac{1}{2} \left(\frac{\pi}{a_{3}}\right)^{1/2} = \frac{\sqrt{\pi}}{2} \left( \frac{2 \Psi_{\lambda}(0)}{\phi_{\lambda}^{(1)}(0)} \right)^{1/2}.
\end{equation}

\paragraph{Oscillatory behavior II}

As a final example, let us consider underdamped Gaussian relaxation. Therefore, we choose the phenomenological ansatz
\begin{equation}
\label{a13}
\Psi_{5,\lambda}(t) \equiv \Psi_{\lambda}(0)\,\e{-a_{5} t^{2}}\, \cos{(b_{5} t)}\,.
\end{equation}
In complete analogy to the previous examples we have
\begin{equation}
\label{a14}
\begin{split}
&\phi_{5,\lambda}(t) = -\frac{\td}{\td t} \Psi_{5,\lambda}(t)\\
&= \Psi_{\lambda}(0) \left[ (2 a_{5} + b_{5}^{2}) t - \left(2 a_{5}^{2} + 2 a_{5} b_{5}^{2} + \frac{b_{5}^{4}}{6}\right) t^{3} + \mc{O}(t^5) \right] \\
&= \phi_{\lambda}^{(1)}(0) t + \phi_{\lambda}^{(3)}(0)\frac{t^{3}}{3!} + \mc{O}(t^5),
\end{split}
\end{equation}
which leads to the following system of equations
\begin{subequations}
\label{a15}
\begin{align}
\phi_{\lambda}^{(1)}(0) &= \Psi_{\lambda}(0) (2 a_{5} + b_{5}^{2}), \\
\phi_{\lambda}^{(3)}(0) &= -6 \Psi_{\lambda}(0) \left(2 a_{5}^{2} + 2 a_{5} b_{5}^{2} + \frac{b_{5}^{4}}{6}\right). 
\end{align}
\end{subequations}
To gain further insight into the physical meaning of Eq.~\eqref{a15} let us consider the Hamiltonian (\ref{b5}) together with Eq.~(\ref{q37}) and the related result (\ref{b11}) for weak coupling, namely $\eta\omega_{D}/\lambda < 1$. In this case, the solution of (\ref{a15}) can be written as
\begin{subequations}
\label{a16}
\begin{align}
a_{5} &= \frac{\lambda}{m}\left[ 1 - \frac{\sqrt{2}}{2}\left(1 - \frac{\eta\omega_{D}}{2\lambda}\right)^{1/2} \right], \\
b_{5} &= 2^{1/4} \left(\frac{\lambda}{m}\right)^{1/2} \left(1 - \frac{\eta\omega_{D}}{2\lambda} \right)^{1/4}.
\end{align}
\end{subequations}

The corresponding correlation time becomes
\begin{equation}
\label{a18}
\tau^{c}_{5} = \sqrt{\frac{\pi m}{2\lambda\,(1-f(\lambda))}} \,
\e{ -\frac{f(\lambda)}{2(1-f(\lambda))}}
\end{equation}
with $f(\lambda) \equiv \sqrt{2 - \eta\omega_{D}/\lambda}/2$.

\section{Obtaining the extrema through variational calculus}
\label{ap3}

In this appendix we show how the extrema of section \ref{sec:example} can be obtained using calculus of variations. The functional 
(\ref{q22}) for $W_{\mathrm{irr}}$ is of the form
\begin{equation}
J[g(s)] = \int_{0}^{1}\td s\, F(g(s),\dot{g}(s)),
\label{c.1}
\end{equation}
where $\dot{g}(s)\equiv \td g/\td s$. The necessary condition for an extrema of (\ref{c.1}) is given by the Euler-Lagrange equation \cite{gelfand}
\begin{equation}
\frac{\td}{\td s}\frac{\partial F}{\partial \dot{g}} - \frac{\partial F}{\partial g} = 0,
\label{c.2}
\end{equation}
together with the fixed end points boundary conditions $g(0) = 0$ and $g(1)=1$.

When $F$ does not depend on $s$ explicitly, (\ref{c.2}) becomes \cite{gelfand}
\begin{equation}
\frac{\td}{\td s}\left( F - \dot{g} \frac{\partial F}{\partial \dot{g}} \right) = 0,
\label{c.4}
\end{equation}
or equivalently
\begin{equation}
F - \dot{g} \frac{\partial F}{\partial\dot{g}} = \mathrm{const}.
\label{c.5}
\end{equation}

If $F$ is also independent on $g(s)$, as in Eqs.~(\ref{q36}) and (\ref{q45}), Eq.~(\ref{c.2}) simply reads
\begin{equation}
\frac{\td}{\td s} \frac{\partial F}{\partial \dot{g}} = 0,
\label{c.6}
\end{equation}
which, for $F = \dot{g}^{2}(s)$, yields $\dot{g}(s) = \mathrm{const}$. Therefore, we finally obtain the extremum 
\begin{equation}
g^{*}(s) = s.
\label{c.8}
\end{equation}

For Eqs.~(\ref{q39}) and (\ref{q49}), $F$ has the form
\begin{equation}
F(g,\dot{g}) = \dot{g}^{2}(s)\left(1 + \mu\, g(s) \right)^{-l},
\label{c.9}
\end{equation}
with $l > 1$ and $\mu > -1$. Equations (\ref{c.5}) and (\ref{c.9}) then yield
\begin{equation}
-\dot{g}^{2}(s) \left( 1+\mu\,g(s) \right)^{-l} = \kappa,
\label{c.10}
\end{equation}
or, equivalently,
\begin{equation}
\dot{g}^{2}(s) + \kappa \left(1 + \mu\,g(s) \right)^{l} = 0,
\label{c.11}
\end{equation}
where $\kappa$ is a constant to be determined by the boundary conditions.

In the case of the harmonic trap, $l = 5/2$ and the solution $g^{*}(s)$ of (\ref{c.11}) is given by
\begin{equation}
g^{*}(s) = -\frac{1}{\mu} + \frac{1}{A (s+B)^{4}},
\label{c.12}
\end{equation}
where $A = \kappa^{2}\mu^{5}/4^{4}$. Demanding (\ref{c.12}) to fulfill the boundary conditions, one obtains
$A^{-1} = B^{4}/\mu$ and
\begin{equation}
B = \left\{ 
\begin{aligned}
\left[ (1 + \mu)^{-1/4} - 1 \right]^{-1},&\;\mathrm{for}\, (1 + \mu) < 1\\
-\left[ 1 - (1 + \mu)^{-1/4} \right]^{-1},&\;\mathrm{for}\, (1 + \mu) > 1. 
\end{aligned}
\right.
\label{c.13}
\end{equation}

In the case of the anharmonic trap, $l = 9/4$ and the solution $g^{*}(s)$ of (\ref{c.11}) is given by
\begin{equation}
g^{*}(s) = -\frac{1}{\mu} + \frac{1}{A (s + B)^{8}},
\label{c.14}
\end{equation}
where $A = \kappa^{4}\mu^{9}/4$. Demanding (\ref{c.14}) to fulfill the boundary conditions, one obtains
$A^{-1} = B^{8}/\mu$ and
\begin{equation}
B = \left\{
\begin{aligned}
\left[ (1 + \mu)^{-1/8} - 1 \right]^{-1},&\;\mathrm{for}\, (1 + \mu) < 1\\
-\left[ 1 - (1 + \mu)^{-1/8} \right]^{-1},&\;\mathrm{for}\, (1 + \mu) > 1. 
\end{aligned}
\right.
\label{c.15}
\end{equation}
%


\begin{thebibliography}{46}%
\makeatletter
\providecommand \@ifxundefined [1]{%
 \@ifx{#1\undefined}
}%
\providecommand \@ifnum [1]{%
 \ifnum #1\expandafter \@firstoftwo
 \else \expandafter \@secondoftwo
 \fi
}%
\providecommand \@ifx [1]{%
 \ifx #1\expandafter \@firstoftwo
 \else \expandafter \@secondoftwo
 \fi
}%
\providecommand \natexlab [1]{#1}%
\providecommand \enquote  [1]{``#1''}%
\providecommand \bibnamefont  [1]{#1}%
\providecommand \bibfnamefont [1]{#1}%
\providecommand \citenamefont [1]{#1}%
\providecommand \href@noop [0]{\@secondoftwo}%
\providecommand \href [0]{\begingroup \@sanitize@url \@href}%
\providecommand \@href[1]{\@@startlink{#1}\@@href}%
\providecommand \@@href[1]{\endgroup#1\@@endlink}%
\providecommand \@sanitize@url [0]{\catcode `\\12\catcode `\$12\catcode
  `\&12\catcode `\#12\catcode `\^12\catcode `\_12\catcode `\%12\relax}%
\providecommand \@@startlink[1]{}%
\providecommand \@@endlink[0]{}%
\providecommand \url  [0]{\begingroup\@sanitize@url \@url }%
\providecommand \@url [1]{\endgroup\@href {#1}{\urlprefix }}%
\providecommand \urlprefix  [0]{URL }%
\providecommand \Eprint [0]{\href }%
\providecommand \doibase [0]{http://dx.doi.org/}%
\providecommand \selectlanguage [0]{\@gobble}%
\providecommand \bibinfo  [0]{\@secondoftwo}%
\providecommand \bibfield  [0]{\@secondoftwo}%
\providecommand \translation [1]{[#1]}%
\providecommand \BibitemOpen [0]{}%
\providecommand \bibitemStop [0]{}%
\providecommand \bibitemNoStop [0]{.\EOS\space}%
\providecommand \EOS [0]{\spacefactor3000\relax}%
\providecommand \BibitemShut  [1]{\csname bibitem#1\endcsname}%
\let\auto@bib@innerbib\@empty
\bibitem [{\citenamefont {Callen}(1985)}]{callen_85}%
  \BibitemOpen
  \bibfield  {author} {\bibinfo {author} {\bibfnamefont {H.}~\bibnamefont
  {Callen}},\ }\href@noop {} {\emph {\bibinfo {title} {Thermodynamics and an
  Introduction to Thermostastistics}}}\ (\bibinfo  {publisher} {Wiley},\
  \bibinfo {address} {New York, USA},\ \bibinfo {year} {1985})\BibitemShut
  {NoStop}%
\bibitem [{\citenamefont {Band}, \citenamefont {Kafri},\ and\ \citenamefont
  {Salamon}(1982)}]{salamon_1982}%
  \BibitemOpen
  \bibfield  {author} {\bibinfo {author} {\bibfnamefont {Y.~B.}\ \bibnamefont
  {Band}}, \bibinfo {author} {\bibfnamefont {O.}~\bibnamefont {Kafri}}, \ and\
  \bibinfo {author} {\bibfnamefont {P.}~\bibnamefont {Salamon}},\ }\href
  {http://scitation.aip.org/content/aip/journal/jap/53/1/10.1063/1.329960}
  {\bibfield  {journal} {\bibinfo  {journal} {J. Appl. Phys.}\ }\textbf
  {\bibinfo {volume} {53}},\ \bibinfo {pages} {8} (\bibinfo {year}
  {1982})}\BibitemShut {NoStop}%
\bibitem [{\citenamefont {Salamon}\ and\ \citenamefont
  {Berry}(1983)}]{salamon_1983}%
  \BibitemOpen
  \bibfield  {author} {\bibinfo {author} {\bibfnamefont {P.}~\bibnamefont
  {Salamon}}\ and\ \bibinfo {author} {\bibfnamefont {R.~S.}\ \bibnamefont
  {Berry}},\ }\href {http://link.aps.org/doi/10.1103/PhysRevLett.51.1127}
  {\bibfield  {journal} {\bibinfo  {journal} {Phys. Rev. Lett.}\ }\textbf
  {\bibinfo {volume} {51}},\ \bibinfo {pages} {1127} (\bibinfo {year}
  {1983})}\BibitemShut {NoStop}%
\bibitem [{\citenamefont {Andresen}, \citenamefont {Salamon},\ and\
  \citenamefont {Berry}(1984)}]{andresen_1984}%
  \BibitemOpen
  \bibfield  {author} {\bibinfo {author} {\bibfnamefont {B.}~\bibnamefont
  {Andresen}}, \bibinfo {author} {\bibfnamefont {P.}~\bibnamefont {Salamon}}, \
  and\ \bibinfo {author} {\bibfnamefont {R.~S.}\ \bibnamefont {Berry}},\ }\href
  {http://link.aip.org/link/PHTOAD/v37/i9/p62/s1&Agg=doi} {\bibfield  {journal}
  {\bibinfo  {journal} {Phys. Today}\ }\textbf {\bibinfo {volume} {37}},\
  \bibinfo {pages} {62} (\bibinfo {year} {1984})}\BibitemShut {NoStop}%
\bibitem [{\citenamefont {Hunter~III.}, \citenamefont {Reinhardt},\ and\
  \citenamefont {Davis}(1993)}]{reinhardt_1993}%
  \BibitemOpen
  \bibfield  {author} {\bibinfo {author} {\bibfnamefont {J.~E.}\ \bibnamefont
  {Hunter~III.}}, \bibinfo {author} {\bibfnamefont {W.~P.}\ \bibnamefont
  {Reinhardt}}, \ and\ \bibinfo {author} {\bibfnamefont {T.~F.}\ \bibnamefont
  {Davis}},\ }\href
  {http://scitation.aip.org/content/aip/journal/jcp/99/9/10.1063/1.465830}
  {\bibfield  {journal} {\bibinfo  {journal} {J. Chem. Phys.}\ }\textbf
  {\bibinfo {volume} {99}},\ \bibinfo {pages} {6856} (\bibinfo {year}
  {1993})}\BibitemShut {NoStop}%
\bibitem [{\citenamefont {de~Koning}\ and\ \citenamefont
  {Antonelli}(1997)}]{dekoning_1997}%
  \BibitemOpen
  \bibfield  {author} {\bibinfo {author} {\bibfnamefont {M.}~\bibnamefont
  {de~Koning}}\ and\ \bibinfo {author} {\bibfnamefont {A.}~\bibnamefont
  {Antonelli}},\ }\href
  {https://journals.aps.org/prb/abstract/10.1103/PhysRevB.55.735} {\bibfield
  {journal} {\bibinfo  {journal} {Phys. Rev. B}\ }\textbf {\bibinfo {volume}
  {55}},\ \bibinfo {pages} {735} (\bibinfo {year} {1997})}\BibitemShut
  {NoStop}%
\bibitem [{\citenamefont {Ytreberg}\ and\ \citenamefont
  {Zuckerman}(2004)}]{zuckerman_2004}%
  \BibitemOpen
  \bibfield  {author} {\bibinfo {author} {\bibfnamefont {F.~M.}\ \bibnamefont
  {Ytreberg}}\ and\ \bibinfo {author} {\bibfnamefont {D.~M.}\ \bibnamefont
  {Zuckerman}},\ }\href
  {http://scitation.aip.org/content/aip/journal/jcp/120/23/10.1063/1.1760511}
  {\bibfield  {journal} {\bibinfo  {journal} {J. Chem. Phys.}\ }\textbf
  {\bibinfo {volume} {120}},\ \bibinfo {pages} {10876} (\bibinfo {year}
  {2004})}\BibitemShut {NoStop}%
\bibitem [{\citenamefont {Geiger}\ and\ \citenamefont
  {Dellago}(2010)}]{dellago_2010}%
  \BibitemOpen
  \bibfield  {author} {\bibinfo {author} {\bibfnamefont {P.}~\bibnamefont
  {Geiger}}\ and\ \bibinfo {author} {\bibfnamefont {C.}~\bibnamefont
  {Dellago}},\ }\href
  {http://journals.aps.org/pre/abstract/10.1103/PhysRevE.81.021127} {\bibfield
  {journal} {\bibinfo  {journal} {Phys. Rev. E}\ }\textbf {\bibinfo {volume}
  {81}},\ \bibinfo {pages} {021127} (\bibinfo {year} {2010})}\BibitemShut
  {NoStop}%
\bibitem [{\citenamefont {Jarzynski}(1997{\natexlab{a}})}]{jarzynski_1997}%
  \BibitemOpen
  \bibfield  {author} {\bibinfo {author} {\bibfnamefont {C.}~\bibnamefont
  {Jarzynski}},\ }\href {http://link.aps.org/doi/10.1103/PhysRevLett.78.2690}
  {\bibfield  {journal} {\bibinfo  {journal} {Phys. Rev. Lett.}\ }\textbf
  {\bibinfo {volume} {78}},\ \bibinfo {pages} {2690} (\bibinfo {year}
  {1997}{\natexlab{a}})}\BibitemShut {NoStop}%
\bibitem [{\citenamefont {Jarzynski}(1997{\natexlab{b}})}]{jarzynski_1997_PRE}%
  \BibitemOpen
  \bibfield  {author} {\bibinfo {author} {\bibfnamefont {C.}~\bibnamefont
  {Jarzynski}},\ }\href {http://link.aps.org/doi/10.1103/PhysRevE.56.5018}
  {\bibfield  {journal} {\bibinfo  {journal} {Phys. Rev. E}\ }\textbf {\bibinfo
  {volume} {56}},\ \bibinfo {pages} {5018} (\bibinfo {year}
  {1997}{\natexlab{b}})}\BibitemShut {NoStop}%
\bibitem [{\citenamefont {Crooks}(1998)}]{crooks_1998}%
  \BibitemOpen
  \bibfield  {author} {\bibinfo {author} {\bibfnamefont {G.~E.}\ \bibnamefont
  {Crooks}},\ }\href
  {http://link.springer.com/article/10.1023%2FA%3A1023208217925?LI=true}
  {\bibfield  {journal} {\bibinfo  {journal} {J. Stat. Phys.}\ }\textbf
  {\bibinfo {volume} {90}},\ \bibinfo {pages} {1481} (\bibinfo {year}
  {1998})}\BibitemShut {NoStop}%
\bibitem [{\citenamefont {Crooks}(1999)}]{crooks_1999}%
  \BibitemOpen
  \bibfield  {author} {\bibinfo {author} {\bibfnamefont {G.~E.}\ \bibnamefont
  {Crooks}},\ }\href {http://link.aps.org/doi/10.1103/PhysRevE.60.2721}
  {\bibfield  {journal} {\bibinfo  {journal} {\PRE}\ }\textbf {\bibinfo
  {volume} {60}},\ \bibinfo {pages} {2721} (\bibinfo {year}
  {1999})}\BibitemShut {NoStop}%
\bibitem [{\citenamefont {Bustamante}\ and\ \citenamefont {and.
  F.~Ritort}(2005)}]{bustamante_2005}%
  \BibitemOpen
  \bibfield  {author} {\bibinfo {author} {\bibfnamefont {C.}~\bibnamefont
  {Bustamante}}\ and\ \bibinfo {author} {\bibfnamefont {J.~L.}\ \bibnamefont
  {and. F.~Ritort}},\ }\href
  {http://link.aip.org/link/PHTOAD/v58/i7/p43/s1&Agg=doi} {\bibfield  {journal}
  {\bibinfo  {journal} {Phys. Today}\ }\textbf {\bibinfo {volume} {58}},\
  \bibinfo {pages} {43} (\bibinfo {year} {2005})}\BibitemShut {NoStop}%
\bibitem [{\citenamefont {Campisi}, \citenamefont {H\"anggi},\ and\
  \citenamefont {Talkner}(2011)}]{campisi_2011}%
  \BibitemOpen
  \bibfield  {author} {\bibinfo {author} {\bibfnamefont {M.}~\bibnamefont
  {Campisi}}, \bibinfo {author} {\bibfnamefont {P.}~\bibnamefont {H\"anggi}}, \
  and\ \bibinfo {author} {\bibfnamefont {P.}~\bibnamefont {Talkner}},\ }\href
  {http://journals.aps.org/rmp/abstract/10.1103/RevModPhys.83.771} {\bibfield
  {journal} {\bibinfo  {journal} {\RMP}\ }\textbf {\bibinfo {volume} {83}},\
  \bibinfo {pages} {771} (\bibinfo {year} {2011})}\BibitemShut {NoStop}%
\bibitem [{\citenamefont {Liphardt}\ \emph {et~al.}(2002)\citenamefont
  {Liphardt}, \citenamefont {Dumont}, \citenamefont {Smith}, \citenamefont
  {Tinoco}, \citenamefont {Jr.},\ and\ \citenamefont
  {Bustamante}}]{liphardt_2002}%
  \BibitemOpen
  \bibfield  {author} {\bibinfo {author} {\bibfnamefont {J.}~\bibnamefont
  {Liphardt}}, \bibinfo {author} {\bibfnamefont {S.}~\bibnamefont {Dumont}},
  \bibinfo {author} {\bibfnamefont {S.~B.}\ \bibnamefont {Smith}}, \bibinfo
  {author} {\bibfnamefont {I.}~\bibnamefont {Tinoco}}, \bibinfo {author}
  {\bibnamefont {Jr.}}, \ and\ \bibinfo {author} {\bibfnamefont
  {C.}~\bibnamefont {Bustamante}},\ }\href
  {http://www.sciencemag.org/content/296/5574/1832.full.html} {\bibfield
  {journal} {\bibinfo  {journal} {Science}\ }\textbf {\bibinfo {volume}
  {296}},\ \bibinfo {pages} {1832} (\bibinfo {year} {2002})}\BibitemShut
  {NoStop}%
\bibitem [{\citenamefont {Collin}\ \emph {et~al.}(2005)\citenamefont {Collin},
  \citenamefont {Ritort}, \citenamefont {Jarzynski}, \citenamefont {Smith},
  \citenamefont {Tinoco},\ and\ \citenamefont {Bustamante}}]{collin_2005}%
  \BibitemOpen
  \bibfield  {author} {\bibinfo {author} {\bibfnamefont {D.}~\bibnamefont
  {Collin}}, \bibinfo {author} {\bibfnamefont {F.}~\bibnamefont {Ritort}},
  \bibinfo {author} {\bibfnamefont {C.}~\bibnamefont {Jarzynski}}, \bibinfo
  {author} {\bibfnamefont {S.~B.}\ \bibnamefont {Smith}}, \bibinfo {author}
  {\bibfnamefont {I.}~\bibnamefont {Tinoco}}, \ and\ \bibinfo {author}
  {\bibfnamefont {C.}~\bibnamefont {Bustamante}},\ }\href
  {http://www.nature.com/nature/journal/v437/n7056/full/nature04061.html}
  {\bibfield  {journal} {\bibinfo  {journal} {Nature (London)}\ }\textbf
  {\bibinfo {volume} {437}},\ \bibinfo {pages} {231} (\bibinfo {year}
  {2005})}\BibitemShut {NoStop}%
\bibitem [{\citenamefont {Zimanyi}\ and\ \citenamefont
  {Silbey}(2009)}]{zimanyi_2009}%
  \BibitemOpen
  \bibfield  {author} {\bibinfo {author} {\bibfnamefont {E.~N.}\ \bibnamefont
  {Zimanyi}}\ and\ \bibinfo {author} {\bibfnamefont {R.~J.}\ \bibnamefont
  {Silbey}},\ }\href
  {http://scitation.aip.org/content/aip/journal/jcp/130/17/10.1063/1.3132747}
  {\bibfield  {journal} {\bibinfo  {journal} {J. Chem. Phys.}\ }\textbf
  {\bibinfo {volume} {130}},\ \bibinfo {pages} {171102} (\bibinfo {year}
  {2009})}\BibitemShut {NoStop}%
\bibitem [{\citenamefont {Andrieux}\ and\ \citenamefont
  {Gaspard}(2004)}]{andrieux_2004}%
  \BibitemOpen
  \bibfield  {author} {\bibinfo {author} {\bibfnamefont {D.}~\bibnamefont
  {Andrieux}}\ and\ \bibinfo {author} {\bibfnamefont {P.}~\bibnamefont
  {Gaspard}},\ }\href
  {http://scitation.aip.org/content/aip/journal/jcp/121/13/10.1063/1.1782391?ver=pdfcov}
  {\bibfield  {journal} {\bibinfo  {journal} {J. Chem. Phys.}\ }\textbf
  {\bibinfo {volume} {121}},\ \bibinfo {pages} {6167} (\bibinfo {year}
  {2004})}\BibitemShut {NoStop}%
\bibitem [{\citenamefont {Andrieux}\ and\ \citenamefont
  {Gaspard}(2008)}]{andrieux_2008}%
  \BibitemOpen
  \bibfield  {author} {\bibinfo {author} {\bibfnamefont {D.}~\bibnamefont
  {Andrieux}}\ and\ \bibinfo {author} {\bibfnamefont {P.}~\bibnamefont
  {Gaspard}},\ }\href {http://link.aps.org/doi/10.1103/PhysRevLett.100.230404}
  {\bibfield  {journal} {\bibinfo  {journal} {\PRL}\ }\textbf {\bibinfo
  {volume} {100}},\ \bibinfo {pages} {230404} (\bibinfo {year}
  {2008})}\BibitemShut {NoStop}%
\bibitem [{\citenamefont {Atilgan}\ and\ \citenamefont {Sun}(2004)}]{sun_2004}%
  \BibitemOpen
  \bibfield  {author} {\bibinfo {author} {\bibfnamefont {E.}~\bibnamefont
  {Atilgan}}\ and\ \bibinfo {author} {\bibfnamefont {S.~X.}\ \bibnamefont
  {Sun}},\ }\href
  {http://scitation.aip.org/content/aip/journal/jcp/121/21/10.1063/1.1813434}
  {\bibfield  {journal} {\bibinfo  {journal} {J. Chem. Phys.}\ }\textbf
  {\bibinfo {volume} {121}},\ \bibinfo {pages} {10392} (\bibinfo {year}
  {2004})}\BibitemShut {NoStop}%
\bibitem [{\citenamefont {Vaikuntanathan}\ and\ \citenamefont
  {Jarzynski}(2008)}]{jarzynski_2008}%
  \BibitemOpen
  \bibfield  {author} {\bibinfo {author} {\bibfnamefont {S.}~\bibnamefont
  {Vaikuntanathan}}\ and\ \bibinfo {author} {\bibfnamefont {C.}~\bibnamefont
  {Jarzynski}},\ }\href
  {http://journals.aps.org/prl/abstract/10.1103/PhysRevLett.100.190601}
  {\bibfield  {journal} {\bibinfo  {journal} {Phys. Rev. Lett.}\ }\textbf
  {\bibinfo {volume} {100}},\ \bibinfo {pages} {190601} (\bibinfo {year}
  {2008})}\BibitemShut {NoStop}%
\bibitem [{\citenamefont {Ballard}\ and\ \citenamefont
  {Jarzynski}(2012)}]{ballard_2012}%
  \BibitemOpen
  \bibfield  {author} {\bibinfo {author} {\bibfnamefont {A.~J.}\ \bibnamefont
  {Ballard}}\ and\ \bibinfo {author} {\bibfnamefont {C.}~\bibnamefont
  {Jarzynski}},\ }\href
  {http://scitation.aip.org/content/aip/journal/jcp/136/19/10.1063/1.4712028}
  {\bibfield  {journal} {\bibinfo  {journal} {J. Chem. Phys.}\ }\textbf
  {\bibinfo {volume} {136}},\ \bibinfo {pages} {194101} (\bibinfo {year}
  {2012})}\BibitemShut {NoStop}%
\bibitem [{\citenamefont {Sivak}, \citenamefont {Chodera},\ and\ \citenamefont
  {Crooks}(2013)}]{sivak_2013}%
  \BibitemOpen
  \bibfield  {author} {\bibinfo {author} {\bibfnamefont {D.~A.}\ \bibnamefont
  {Sivak}}, \bibinfo {author} {\bibfnamefont {J.~D.}\ \bibnamefont {Chodera}},
  \ and\ \bibinfo {author} {\bibfnamefont {G.~E.}\ \bibnamefont {Crooks}},\
  }\href {http://link.aps.org/doi/10.1103/PhysRevX.3.011007} {\bibfield
  {journal} {\bibinfo  {journal} {Phys. Rev. X}\ }\textbf {\bibinfo {volume}
  {3}},\ \bibinfo {pages} {011007} (\bibinfo {year} {2013})}\BibitemShut
  {NoStop}%
\bibitem [{\citenamefont {Schmiedl}\ and\ \citenamefont
  {Seifert}(2007)}]{seifert_2007}%
  \BibitemOpen
  \bibfield  {author} {\bibinfo {author} {\bibfnamefont {T.}~\bibnamefont
  {Schmiedl}}\ and\ \bibinfo {author} {\bibfnamefont {U.}~\bibnamefont
  {Seifert}},\ }\href
  {http://journals.aps.org/prl/abstract/10.1103/PhysRevLett.98.108301}
  {\bibfield  {journal} {\bibinfo  {journal} {Phys. Rev Lett.}\ }\textbf
  {\bibinfo {volume} {98}},\ \bibinfo {pages} {108301} (\bibinfo {year}
  {2007})}\BibitemShut {NoStop}%
\bibitem [{\citenamefont {Gomez-Marin}, \citenamefont {Schmiedl},\ and\
  \citenamefont {Seifert}(2008)}]{seifert_2008}%
  \BibitemOpen
  \bibfield  {author} {\bibinfo {author} {\bibfnamefont {A.}~\bibnamefont
  {Gomez-Marin}}, \bibinfo {author} {\bibfnamefont {T.}~\bibnamefont
  {Schmiedl}}, \ and\ \bibinfo {author} {\bibfnamefont {U.}~\bibnamefont
  {Seifert}},\ }\href
  {http://scitation.aip.org/content/aip/journal/jcp/129/2/10.1063/1.2948948}
  {\bibfield  {journal} {\bibinfo  {journal} {J. Chem. Phys.}\ }\textbf
  {\bibinfo {volume} {129}},\ \bibinfo {pages} {024114} (\bibinfo {year}
  {2008})}\BibitemShut {NoStop}%
\bibitem [{\citenamefont {Then}\ and\ \citenamefont {Engel}(2008)}]{then_2008}%
  \BibitemOpen
  \bibfield  {author} {\bibinfo {author} {\bibfnamefont {H.}~\bibnamefont
  {Then}}\ and\ \bibinfo {author} {\bibfnamefont {A.}~\bibnamefont {Engel}},\
  }\href {http://journals.aps.org/pre/abstract/10.1103/PhysRevE.77.041105}
  {\bibfield  {journal} {\bibinfo  {journal} {Phys. Rev. E}\ }\textbf {\bibinfo
  {volume} {77}},\ \bibinfo {pages} {041105} (\bibinfo {year}
  {2008})}\BibitemShut {NoStop}%
\bibitem [{\citenamefont {Aurell}, \citenamefont {Mej\'ia-Monasterio},\ and\
  \citenamefont {Muratore-Ginanneschi}(2011)}]{aurell_2011}%
  \BibitemOpen
  \bibfield  {author} {\bibinfo {author} {\bibfnamefont {E.}~\bibnamefont
  {Aurell}}, \bibinfo {author} {\bibfnamefont {C.}~\bibnamefont
  {Mej\'ia-Monasterio}}, \ and\ \bibinfo {author} {\bibfnamefont
  {P.}~\bibnamefont {Muratore-Ginanneschi}},\ }\href
  {http://journals.aps.org/prl/abstract/10.1103/PhysRevLett.106.250601}
  {\bibfield  {journal} {\bibinfo  {journal} {Phys. Rev. Lett.}\ }\textbf
  {\bibinfo {volume} {106}},\ \bibinfo {pages} {250601} (\bibinfo {year}
  {2011})}\BibitemShut {NoStop}%
\bibitem [{\citenamefont {de~Koning}(2005)}]{dekoning_2005}%
  \BibitemOpen
  \bibfield  {author} {\bibinfo {author} {\bibfnamefont {M.}~\bibnamefont
  {de~Koning}},\ }\href
  {http://scitation.aip.org/content/aip/journal/jcp/122/10/10.1063/1.1860556}
  {\bibfield  {journal} {\bibinfo  {journal} {J. Chem. Phys}\ }\textbf
  {\bibinfo {volume} {122}},\ \bibinfo {pages} {104106} (\bibinfo {year}
  {2005})}\BibitemShut {NoStop}%
\bibitem [{\citenamefont {Lindberg}, \citenamefont {Berkelbach},\ and\
  \citenamefont {Wang}(2009)}]{lindberg_2009}%
  \BibitemOpen
  \bibfield  {author} {\bibinfo {author} {\bibfnamefont {G.~E.}\ \bibnamefont
  {Lindberg}}, \bibinfo {author} {\bibfnamefont {T.~C.}\ \bibnamefont
  {Berkelbach}}, \ and\ \bibinfo {author} {\bibfnamefont {F.}~\bibnamefont
  {Wang}},\ }\href
  {http://scitation.aip.org/content/aip/journal/jcp/130/17/10.1063/1.3126602}
  {\bibfield  {journal} {\bibinfo  {journal} {J. Chem. Phys.}\ }\textbf
  {\bibinfo {volume} {130}},\ \bibinfo {pages} {174705} (\bibinfo {year}
  {2009})}\BibitemShut {NoStop}%
\bibitem [{\citenamefont {Sivak}\ and\ \citenamefont
  {Crooks}(2012)}]{crooks_2012}%
  \BibitemOpen
  \bibfield  {author} {\bibinfo {author} {\bibfnamefont {D.~A.}\ \bibnamefont
  {Sivak}}\ and\ \bibinfo {author} {\bibfnamefont {G.~E.}\ \bibnamefont
  {Crooks}},\ }\href {http://link.aps.org/doi/10.1103/PhysRevLett.108.190602}
  {\bibfield  {journal} {\bibinfo  {journal} {Phys. Rev. Lett.}\ }\textbf
  {\bibinfo {volume} {108}},\ \bibinfo {pages} {190602} (\bibinfo {year}
  {2012})}\BibitemShut {NoStop}%
\bibitem [{\citenamefont {Zulkowski}\ \emph {et~al.}(2012)\citenamefont
  {Zulkowski}, \citenamefont {Sivak}, \citenamefont {Crooks},\ and\
  \citenamefont {DeWeese}}]{crooks_2012_PRE}%
  \BibitemOpen
  \bibfield  {author} {\bibinfo {author} {\bibfnamefont {P.~R.}\ \bibnamefont
  {Zulkowski}}, \bibinfo {author} {\bibfnamefont {D.~A.}\ \bibnamefont
  {Sivak}}, \bibinfo {author} {\bibfnamefont {G.~E.}\ \bibnamefont {Crooks}}, \
  and\ \bibinfo {author} {\bibfnamefont {M.~R.}\ \bibnamefont {DeWeese}},\
  }\href {http://link.aps.org/doi/10.1103/PhysRevE.86.041148} {\bibfield
  {journal} {\bibinfo  {journal} {Phys. Rev. E}\ }\textbf {\bibinfo {volume}
  {86}},\ \bibinfo {pages} {041148} (\bibinfo {year} {2012})}\BibitemShut
  {NoStop}%
\bibitem [{\citenamefont {Campisi}, \citenamefont {Denisov},\ and\
  \citenamefont {H\"anggi}(2012)}]{campisi_2012}%
  \BibitemOpen
  \bibfield  {author} {\bibinfo {author} {\bibfnamefont {M.}~\bibnamefont
  {Campisi}}, \bibinfo {author} {\bibfnamefont {S.}~\bibnamefont {Denisov}}, \
  and\ \bibinfo {author} {\bibfnamefont {P.}~\bibnamefont {H\"anggi}},\ }\href
  {\doibase 10.1103/PhysRevA.86.032114} {\bibfield  {journal} {\bibinfo
  {journal} {Phys. Rev. A}\ }\textbf {\bibinfo {volume} {86}},\ \bibinfo
  {pages} {032114} (\bibinfo {year} {2012})}\BibitemShut {NoStop}%
\bibitem [{\citenamefont {{Thingna}}\ \emph {et~al.}(2014)\citenamefont
  {{Thingna}}, \citenamefont {{H{\"a}nggi}}, \citenamefont {{Fazio}},\ and\
  \citenamefont {{Campisi}}}]{thinga_2014}%
  \BibitemOpen
  \bibfield  {author} {\bibinfo {author} {\bibfnamefont {J.}~\bibnamefont
  {{Thingna}}}, \bibinfo {author} {\bibfnamefont {P.}~\bibnamefont
  {{H{\"a}nggi}}}, \bibinfo {author} {\bibfnamefont {R.}~\bibnamefont
  {{Fazio}}}, \ and\ \bibinfo {author} {\bibfnamefont {M.}~\bibnamefont
  {{Campisi}}},\ }\href {http://arxiv.org/abs/1403.3523} {\bibfield  {journal}
  {\bibinfo  {journal} {arXiv:1403.3523}\ } (\bibinfo {year}
  {2014})}\BibitemShut {NoStop}%
\bibitem [{\citenamefont {Chernyak}, \citenamefont {Chertkov},\ and\
  \citenamefont {Jarzynski}(2006)}]{chernyak_2005}%
  \BibitemOpen
  \bibfield  {author} {\bibinfo {author} {\bibfnamefont {V.~Y.}\ \bibnamefont
  {Chernyak}}, \bibinfo {author} {\bibfnamefont {M.}~\bibnamefont {Chertkov}},
  \ and\ \bibinfo {author} {\bibfnamefont {C.}~\bibnamefont {Jarzynski}},\
  }\href {http://stacks.iop.org/1742-5468/2006/i=08/a=P08001} {\bibfield
  {journal} {\bibinfo  {journal} {J. Stat. Mech.}\ }\textbf {\bibinfo {volume}
  {2006}},\ \bibinfo {pages} {P08001} (\bibinfo {year} {2006})}\BibitemShut
  {NoStop}%
\bibitem [{\citenamefont {Deffner}, \citenamefont {Brunner},\ and\
  \citenamefont {Lutz}(2011)}]{deffner_2011}%
  \BibitemOpen
  \bibfield  {author} {\bibinfo {author} {\bibfnamefont {S.}~\bibnamefont
  {Deffner}}, \bibinfo {author} {\bibfnamefont {M.}~\bibnamefont {Brunner}}, \
  and\ \bibinfo {author} {\bibfnamefont {E.}~\bibnamefont {Lutz}},\ }\href
  {http://iopscience.iop.org/0295-5075/94/3/30001/} {\bibfield  {journal}
  {\bibinfo  {journal} {\EPL}\ }\textbf {\bibinfo {volume} {94}},\ \bibinfo
  {pages} {30001} (\bibinfo {year} {2011})}\BibitemShut {NoStop}%
\bibitem [{\citenamefont {Vaikuntanathan}\ and\ \citenamefont
  {Jarzynski}(2009)}]{suri_2009}%
  \BibitemOpen
  \bibfield  {author} {\bibinfo {author} {\bibfnamefont {S.}~\bibnamefont
  {Vaikuntanathan}}\ and\ \bibinfo {author} {\bibfnamefont {C.}~\bibnamefont
  {Jarzynski}},\ }\href {http://stacks.iop.org/0295-5075/87/i=6/a=60005}
  {\bibfield  {journal} {\bibinfo  {journal} {EPL (Europhysics Letters)}\
  }\textbf {\bibinfo {volume} {87}},\ \bibinfo {pages} {60005} (\bibinfo {year}
  {2009})}\BibitemShut {NoStop}%
\bibitem [{\citenamefont {Kubo}(1957)}]{kubo_1957}%
  \BibitemOpen
  \bibfield  {author} {\bibinfo {author} {\bibfnamefont {R.}~\bibnamefont
  {Kubo}},\ }\href {http://journals.jps.jp/doi/abs/10.1143/JPSJ.12.570}
  {\bibfield  {journal} {\bibinfo  {journal} {J. Phys. Soc. Jpn.}\ }\textbf
  {\bibinfo {volume} {12}},\ \bibinfo {pages} {570} (\bibinfo {year}
  {1957})}\BibitemShut {NoStop}%
\bibitem [{\citenamefont {Kubo}, \citenamefont {Toda},\ and\ \citenamefont
  {Hashitsume}(1985)}]{kubo_1985}%
  \BibitemOpen
  \bibfield  {author} {\bibinfo {author} {\bibfnamefont {R.}~\bibnamefont
  {Kubo}}, \bibinfo {author} {\bibfnamefont {M.}~\bibnamefont {Toda}}, \ and\
  \bibinfo {author} {\bibfnamefont {N.}~\bibnamefont {Hashitsume}},\
  }\href@noop {} {\emph {\bibinfo {title} {Statistical Physics II}}}\ (\bibinfo
   {publisher} {Springer-Verlag},\ \bibinfo {address} {Berlin},\ \bibinfo
  {year} {1985})\BibitemShut {NoStop}%
\bibitem [{\citenamefont {Nulton}\ \emph {et~al.}(1985)\citenamefont {Nulton},
  \citenamefont {Salamon}, \citenamefont {Andresen},\ and\ \citenamefont
  {Anmin}}]{nulton_1985}%
  \BibitemOpen
  \bibfield  {author} {\bibinfo {author} {\bibfnamefont {J.}~\bibnamefont
  {Nulton}}, \bibinfo {author} {\bibfnamefont {P.}~\bibnamefont {Salamon}},
  \bibinfo {author} {\bibfnamefont {B.}~\bibnamefont {Andresen}}, \ and\
  \bibinfo {author} {\bibfnamefont {Q.}~\bibnamefont {Anmin}},\ }\href
  {http://scitation.aip.org/content/aip/journal/jcp/83/1/10.1063/1.449774}
  {\bibfield  {journal} {\bibinfo  {journal} {J. Chem. Phys.}\ }\textbf
  {\bibinfo {volume} {83}},\ \bibinfo {pages} {334} (\bibinfo {year}
  {1985})}\BibitemShut {NoStop}%
\bibitem [{\citenamefont {Deffner}\ and\ \citenamefont
  {Jarzynski}(2013)}]{deffner_jarzynski_2013}%
  \BibitemOpen
  \bibfield  {author} {\bibinfo {author} {\bibfnamefont {S.}~\bibnamefont
  {Deffner}}\ and\ \bibinfo {author} {\bibfnamefont {C.}~\bibnamefont
  {Jarzynski}},\ }\href
  {http://journals.aps.org/prx/abstract/10.1103/PhysRevX.3.041003} {\bibfield
  {journal} {\bibinfo  {journal} {\PRX}\ }\textbf {\bibinfo {volume} {3}},\
  \bibinfo {pages} {041003} (\bibinfo {year} {2013})}\BibitemShut {NoStop}%
\bibitem [{\citenamefont {Tsao}, \citenamefont {Sheu},\ and\ \citenamefont
  {Mou}(1994)}]{tsao_1994}%
  \BibitemOpen
  \bibfield  {author} {\bibinfo {author} {\bibfnamefont {L.~W.}\ \bibnamefont
  {Tsao}}, \bibinfo {author} {\bibfnamefont {S.~Y.}\ \bibnamefont {Sheu}}, \
  and\ \bibinfo {author} {\bibfnamefont {C.~Y.}\ \bibnamefont {Mou}},\ }\href
  {http://scitation.aip.org/content/aip/journal/jcp/101/3/10.1063/1.467670}
  {\bibfield  {journal} {\bibinfo  {journal} {J. Chem. Phys.}\ }\textbf
  {\bibinfo {volume} {101}},\ \bibinfo {pages} {2302} (\bibinfo {year}
  {1994})}\BibitemShut {NoStop}%
\bibitem [{\citenamefont {Gelfand}\ and\ \citenamefont
  {Fomin}(2000)}]{gelfand}%
  \BibitemOpen
  \bibfield  {author} {\bibinfo {author} {\bibfnamefont {I.~M.}\ \bibnamefont
  {Gelfand}}\ and\ \bibinfo {author} {\bibfnamefont {S.~V.}\ \bibnamefont
  {Fomin}},\ }\href@noop {} {\emph {\bibinfo {title} {Calculus of
  Variations}}}\ (\bibinfo  {publisher} {Dover Publications},\ \bibinfo
  {address} {New York},\ \bibinfo {year} {2000})\BibitemShut {NoStop}%
\bibitem [{\citenamefont {Deffner}\ and\ \citenamefont
  {Bonan\c{c}a}(2014)}]{deffner_2014_letter}%
  \BibitemOpen
  \bibfield  {author} {\bibinfo {author} {\bibfnamefont {S.}~\bibnamefont
  {Deffner}}\ and\ \bibinfo {author} {\bibfnamefont {M.~V.~S.}\ \bibnamefont
  {Bonan\c{c}a}},\ }\href@noop {} {\bibfield  {journal} {\bibinfo  {journal}
  {to be published}\ } (\bibinfo {year} {2014})}\BibitemShut {NoStop}%
\bibitem [{\citenamefont {van Kampen}(2007)}]{vankampen_2007}%
  \BibitemOpen
  \bibfield  {author} {\bibinfo {author} {\bibfnamefont {N.~G.}\ \bibnamefont
  {van Kampen}},\ }\href@noop {} {\emph {\bibinfo {title} {Stochastic Processes
  in Physics and Chemistry}}}\ (\bibinfo  {publisher} {North-Holland},\
  \bibinfo {address} {Amsterdam},\ \bibinfo {year} {2007})\BibitemShut
  {NoStop}%
\bibitem [{\citenamefont {Kubo}\ and\ \citenamefont
  {Ichimura}(1972)}]{kubo_1972}%
  \BibitemOpen
  \bibfield  {author} {\bibinfo {author} {\bibfnamefont {R.}~\bibnamefont
  {Kubo}}\ and\ \bibinfo {author} {\bibfnamefont {M.}~\bibnamefont
  {Ichimura}},\ }\href
  {http://scitation.aip.org/content/aip/journal/jmp/13/10/10.1063/1.1665862}
  {\bibfield  {journal} {\bibinfo  {journal} {J. Math. Phys.}\ }\textbf
  {\bibinfo {volume} {13}},\ \bibinfo {pages} {1454} (\bibinfo {year}
  {1972})}\BibitemShut {NoStop}%
\bibitem [{\citenamefont {Zwanzig}(2001)}]{zwanzig_2001}%
  \BibitemOpen
  \bibfield  {author} {\bibinfo {author} {\bibfnamefont {R.}~\bibnamefont
  {Zwanzig}},\ }\href@noop {} {\emph {\bibinfo {title} {Nonequilibrium
  Statistical Mechanics}}}\ (\bibinfo  {publisher} {Oxford University Press},\
  \bibinfo {address} {Oxford, UK},\ \bibinfo {year} {2001})\BibitemShut
  {NoStop}%
\end{thebibliography}

%

\end{document}